\documentclass[epj]{svjour}

\usepackage[pdftex]{graphicx}
\usepackage[english]{babel}
\usepackage{latexsym,amsmath,verbatim}
\usepackage{color}
\usepackage{microtype}
\usepackage{hyperref}

\newcommand{\av}[1]{\left\langle {#1} \right\rangle}
\newcommand{\be}{\begin{equation}}
\newcommand{\ee}{\end{equation}}

\graphicspath{{./}}

\begin{document}

\title{On the numerical study of percolation and epidemic critical
  properties in networks}

\author{Claudio Castellano\inst{1,2} \and Romualdo Pastor-Satorras\inst{3}} 

\institute{Istituto dei Sistemi Complessi (ISC-CNR), Via dei Taurini 19,
  I-00185 Roma, Italy \and Dipartimento di Fisica, ``Sapienza''
  Universit\`a di Roma, P.le A. Moro 2, I-00185 Roma, Italy \and
  Departament de F\'\i sica, Universitat Polit\`ecnica de Catalunya,
  Campus Nord B4, 08034 Barcelona, Spain}

\titlerunning{Numerical study of percolation and epidemics critical
  properties in networks}

\authorrunning{C. Castellano and R. Pastor-Satorras}

\date{Received: date / Revised version: date}

\abstract{%
  The static properties of the fundamental model for epidemics of
  diseases allowing immunity (susceptible-infected-removed model) are
  known to be derivable by an exact mapping to bond percolation.  Yet
  when performing numerical simulations of these dynamics in a network a
  number of subtleties must be taken into account in order to correctly
  estimate the transition point and the associated critical properties.
  We expose these subtleties and identify the different quantities which
  play the role of criticality detector in the two dynamics.
} 

\maketitle

\section{Introduction}
\label{sec:introduction}

Epidemic processes on complex heterogeneous topologies, such as those
representing social contact networks~\cite{Jackson2010}, can exhibit
surprising features when compared with regular or fully mixed substrates
\cite{Pastor-Satorras:2014aa}. Particular among those hallmarks is a
vanishing epidemic threshold~\cite{pv01a,Castellano2010}, which makes
heterogeneous networks exceedingly prone to the spreading of an
infection, even in the case of a very small infective power. A
cornerstone model for the understanding of diseases that confer immunity
is the susceptible-infected-removed (SIR)
model~\cite{Diekmann_Heesterbeek_Britton_boek2012}. In this model, the
nodes in the network (individuals) can be in three different states:
susceptible ($S$), i.e. able to contract the disease; infected ($I$),
i.e.  able to propagate the disease to a nearest neighbor contact; and
removed ($R$), immune to the disease. The dynamics of the model is as
follows: each infected node connected to a susceptible node can
propagate the disease to the latter with a rate (probability per unit
time) $\lambda$; on the other hand, each infected individual recovers
and becomes removed with a rate $\mu$ (which, without loss of
generality, we fix to $\mu = 1$). Notice that the total rate of
infection of neighbours of a node is proportional to the number of 
susceptible ones.
The behavior of the SIR model is
characterized in terms of the statistical properties of the epidemic
outbreaks it generates, measured by the average number of removed
individuals $N_R$ at the end of an outbreak. In this sense, it is
important to discern the existence of an epidemic threshold $\lambda_c$,
separating a phase $\lambda \leq \lambda_c$ in which the total number of
affected individuals is sublinear with the network size $N$, with
$N_R/N \to 0$ for $N \to \infty$, from a phase $\lambda > \lambda_c$ in
which the disease affects a finite fraction of the population,
$N_R/N \to \mathrm{const.}>0$ for $N \to \infty$.

The properties of the SIR model have been analytically studied applying
different approaches~\cite{Pastor-Satorras:2014aa}.  In particular the
so-called heterogeneous mean-field theory
(HMF)~\cite{refId0,lloyd01,marianproc} focuses on the dynamic properties
of nodes grouped in classes with the same degree and assumes an annealed
network approximation~\cite{dorogovtsev07:_critic_phenom}, neglecting
the actual network structure and considering only an ensemble of random
networks, all sharing some statistical properties (degree distribution,
degree correlations)~\cite{Boguna09}.  An alternative, more accurate,
approach is based on a mapping of SIR outbreaks to a bond percolation
process in the network, with a percolation probability depending on the
rate of infection and recovery of the SIR
process~\cite{newman02,Pastor-Satorras:2014aa}.  Both approaches predict
the presence of an epidemic threshold that, in degree uncorrelated
networks \cite{Dorogovtsev:2002}, is a function of the first $\av{k}$
and second $\av{k^2}$ moment of the network's degree distribution.

Despite the strength of these theoretical predictions the use of
numerical techniques is still important, for both percolation and
SIR on networks, since the theoretical approaches are based on the
omission of topological and dynamical
correlations~\cite{Dorogovtsev:2002}.  Most computational efforts
devoted to determine the position of the critical point in either SIR or
percolation
\cite{Lagorio2009755,Colomer-de-Simon2014,PhysRevE.91.010801,Shu15} rely
on some form of numerical ``susceptibility'', defined in terms of
moments of the cluster or outbreak sizes.  These susceptibilities are
designed to behave as ``criticality detectors'', in the sense that they
should exhibit a peak in the vicinity of the critical point, and
decrease sensibly away from it. The variation of the position and height
of the peak as a function of the network size is then used to determine
the position of the critical point in the thermodynamic (infinite
network) limit, as well as the associated critical exponents, by
applying finite-size scaling (FSS) theory \cite{cardy88}. While the
susceptibilities used so far in the literature usually work numerically,
there has not been, to the best of our knowledge, any effort to put them
on a sound theoretical footing, in particular, in what refers to the
differences in the susceptibilities used for percolation and the SIR
model, and in the relation with the mapping from one process to the
other.

The purpose of this paper is to undertake this effort, by examining
critically the different numerical methods applied so far to determine
the critical properties of the bond percolation and the SIR transitions.
We find that, despite the exact mapping existing between the two models,
different quantities must be considered in the numerical study of
the two cases.  Moreover, particular care must be used in the evaluation
of the critical properties of SIR and its comparison with theoretical
predictions.  

The paper is structured as follows. In Sec.~\ref{sec:mapp-perc-sir} we
briefly summarize the results for percolation and SIR that will be
needed in the rest of the paper.  In Sec.~\ref{sec:numer-analys-perc} we
define different quantities that may be detectors of criticality for
percolation processes on networks, we determine analytically and verify
numerically which one is the most suitable.  We then adapt the same
framework to the SIR model, highlighting the differences between the two
cases, which result in different quantities being optimal detectors of
criticality.  A discussion of the implications of our findings and a
reinterpretation of previous literature, followed by conclusions
is presented in Sec.~\ref{conclusions}.

\section{Background on percolation and the SIR model}
\label{sec:mapp-perc-sir}  

In this section we briefly summarize standard results
on percolation and the SIR model that will be needed 
in the rest of the paper.

\subsection{Percolation in lattices}
\label{sec:percolation-networks}

In bond percolation, edges are removed with probability $1-p$ and kept
with probability $p$.  In regular Euclidean lattices \cite{stauffer94},
a critical value $p_c$ separates a subcritical phase at $p \leq p_c$ in
which only small clusters of connected sites exist, from a supercritical
phase at $p > p_c$, where there is an infinite, spanning cluster. In
finite Euclidean systems, the spanning cluster is defined as any cluster
that touches two opposite boundaries along a given coordinate axis. The
order parameter is thus defined as the probability $\mathcal{P}$ (the
percolation strength) that a randomly selected site belongs to the
spanning cluster.  In an infinite system, above the transition
\begin{equation}
  \mathcal{P} \sim (p-p_c)^{\beta},
  \label{eq:23}
\end{equation}
defining the critical exponent $\beta$. The principal quantity in
percolation, from which all others can be derived, is the normalized
cluster number, $n_s(p)$, defined as the number of finite clusters of
size $s$ per lattice site.  A consequence of this definition is that
$s n_s(p) / \sum_{s'} s' n_{s'}(p)$ is the probability that a randomly
chosen site belongs to a finite cluster of size $s$ \cite{stauffer94}.
In order to determine numerically both the critical point and the
associated critical exponents, one usually studies the so-called mean
cluster size (or susceptibility) \cite{stauffer94},
\begin{equation}
\chi = \sum_s  s \frac{s n_s(p)}{\sum_{s'} s' n_{s'}(p)}.
  \label{eq:16}
\end{equation}
The susceptibility $\chi$ assumes, in an infinite lattice, a finite value 
for any $p$ except at the critical point, $p_c$, where it diverges 
as
\begin{equation} 
  \chi \sim |p-p_c|^{-\gamma}
  \label{eq:24}
\end{equation}
This divergence is related to the scaling form of the normalized cluster
number, which, close to the critical point, obeys
\begin{equation}
  n_s(p) \simeq s^{-\tau} {\cal F}(s \Delta^{1/\sigma}),
  \label{eq:4}
\end{equation}
with $\Delta = |p - p_c|$, $\cal{F}$ is a scaling function, 
while $\tau$ and $\sigma$ are other critical
exponents, related to $\beta$ and $\gamma$ through the
relations~\cite{stauffer94}
\begin{equation}
\sigma = \frac{1}{\beta+\gamma}, \quad \tau = 3 - \frac{\gamma}{\beta +
  \gamma} .
\label{betagamma}
\end{equation}

Right at the critical point, we can apply FSS theory
\cite{stauffer94,cardy88} to see how quantities depend on system size in
finite systems. The basic FSS hypothesis states that the system size
dependence enters in the system by the ratio $\xi / L$, where $L$ is the
longitudinal length and $\xi \sim \Delta^{-\nu}$ the correlation length
\cite{stauffer94}.
Thus, assuming that the order parameter follows, close to criticality,
the scaling form
\begin{equation}
  \mathcal{P}(p, L) = L^{-\beta/\nu} F(\Delta^\nu L)
  \label{eq:15}
\end{equation}
we are led to susceptibility scaling at the critical point as 
\begin{equation}
  \chi(p_c) \simeq  L^{\gamma/\nu},
  \label{eq:14}
\end{equation}
while the order parameter scales as
\begin{equation}
  \mathcal{P}(p_c) \simeq L^{-\beta/\nu}.
\end{equation}
Notice that these last expressions can be simply obtained by replacing
$\Delta \sim L^{-1/\nu}$ in Eqs.~(\ref{eq:23}) and~(\ref{eq:24}).

\subsection{Percolation in networks}
\label{sec:percolation-networks-1}

In networks 
all definitions presented above can be used, with only one caveat: Since
there is no network boundary, it is not possible to define a spanning
cluster, and hence one has to use alternative definitions of the order
parameter $\mathcal{P}$.  The natural modification involves the
consideration of the largest component of the network, which has size
$S$. In the limit of infinite network, below the critical point the size
of the largest component is subextensive: $S/N \to 0$; above the
critical point, on the other hand, the largest component is the giant
connected component of the network \cite{Newman10}, with a size
$S \equiv G \sim N$, proportional to the network size. In this way, we
can define the percolation strength as
$\mathcal{P} = G/N$, which is finite above the critical point.

For random uncorrelated networks the condition for the existence of 
the giant component is~\cite{Callaway2000}
\begin{equation}
  p > p_c = \frac{\av{k}}{\av{k^2} - \av{k}}.
  \label{eq:1}
\end{equation}
The behavior of the order parameter close to criticality is
${\cal P} \sim (p - p_c)^{\beta}$, where the critical exponent
$\beta$ depends on the form of the degree distribution.  For
scale-free networks with a degree distribution
$P(k) \sim k^{-\gamma_d}$, one finds~\cite{Cohen02}
\begin{equation}
  \beta = \left \{
    \begin{array}{ccc}
      1/(3-\gamma_d) & \mathrm{for}& 2< \gamma_d<3 \\
      1/(\gamma_d - 3) & \mathrm{for}& 3< \gamma_d<4\\
      1 & \mathrm{for}& \gamma_d> 4\
    \end{array} \right. .
  \label{eq:9}
\end{equation}
Applying the FSS theory, and assuming a scaling of the order parameter,
following Eq.~(\ref{eq:15}), in the form
\begin{equation}
  \mathcal{P}(p, N) = N^{-\beta/\nu} F(\Delta^\nu N),
  \label{eq:18}
\end{equation}
one finds~\cite{Radicchi15,Wu07}
\begin{equation}
  \nu = \left \{
    \begin{array}{ccc}
      2/(3-\gamma_d) & \mathrm{for}& 2< \gamma_d<3 \\
      (\gamma_d - 1)/(\gamma_d - 3) & \mathrm{for}& 3< \gamma_d<4\\
      3 & \mathrm{for}& \gamma_d> 4\
    \end{array} \right. .
  \label{eq:19}
\end{equation}
Notice that $\nu$ describes the scaling with respect to the network size
and, for the case $2< \gamma_d<3$, one assumes a maximum degree in the
network scaling as $k_\mathrm{max} \sim  N^{1/2}$~\cite{mariancutofss}.
For any $\gamma_d > 3$~\cite{Cohen02}, it holds 
\begin{equation}
\gamma=1,
\label{eq:220}
\end{equation}
while for $2<\gamma_d<3$ the exponent $\gamma$ is effectively 
0~\cite{Radicchi15}.

\subsection{Mapping SIR to percolation}
\label{sec:mapp-sir-perc}

The possibility of studying the SIR model by mapping it to a percolation
process was observed as early as in
Refs.~\cite{Ludwig1975,Grassberger1983}.  In networks, the mapping is
worked out as follows~\cite{newman02}.  Let us consider a modified SIR
model, in which infected nodes remain in this state for a fixed time
$\tau$ after infection. Consider now an infected node and an edge
joining it to a susceptible node. During the infection time $\tau$,
since the transmission of the disease along the edge follows a Poisson
process with rate $\lambda$, the probability that the infection will be
transmitted along the edge is given by the transmissibility $T_\tau$,
which takes the value \cite{renewal}
\begin{equation}
  T_\tau = 1 - e^{-\lambda \tau}.
  \label{eq:10}
\end{equation}
As this transmissibility is the same for all infected nodes and
edges, it is clear that the set of removed nodes generated by a
SIR outbreak starting from a single infected node will be equal to
the connected cluster the initial infected node belongs to 
in a bond percolation process with occupation probability $p = T_\tau$.
From this mapping, the presence of a critical occupation probability $p_c$
implies the existence of a critical transmissibility $T_{\tau, c}$, 
which translates into a critical spreading rate $\lambda_c$.
For uncorrelated networks, Eq.~(\ref{eq:1}) for $p_c$ implies,
using Eq.~(\ref{eq:10}),
\begin{equation}
  \lambda_c = \frac{1}{\tau}\ln \frac{\av{k^2} -  \av{k}}{\av{k^2} -
    2\av{k}}.
  \label{eq:11}
\end{equation}
The previous expression was derived assuming a constant infection time
$\tau$. In general, the original definition of the SIR model, in terms
of a constant recovery rate $\mu$, implies that recovery is a
Poisson process, with a distribution of recovery times
$P_\mathrm{rec}(\tau) = \mu e^{-\tau \mu}$~\cite{renewal}. One
possibility to deal with this fact is to consider the average
transmissibility
\begin{equation}
  \av{T} = \int_0^\infty T_\tau P_\mathrm{rec}(\tau) \; d\tau =
  \frac{\lambda}{1 + \lambda},
 \label{eq:117}
\end{equation}
where we have set $\mu=1$~\cite{newman02}. 
The averaging performed in Eq.~(\ref{eq:117}) is in principle
an approximation, which nevertheless leads to exact results for
the threshold~\cite{Kenah2007,Pastor-Satorras:2014aa}.
In the case of uncorrelated networks, using Eq.~(\ref{eq:1}), 
the exact threshold is
\begin{equation}
  \lambda_c = \frac{\av{k}}{\av{k^2} - 2\av{k}}.
  \label{eq:12}
\end{equation}

\section{Numerical analysis of percolation and the SIR model on
  networks}
\label{sec:numer-analys-perc} 

\subsection{Percolation}
\label{sec:percolation-analysis}

From a numerical point of view, the identification of the percolation
critical point in regular lattices can be performed by applying the FSS
hypothesis to the susceptibility $\chi$. Thus, assuming the analogous
scaling form \cite{stauffer94}
\begin{equation}
  \chi(p, L) \simeq L^{\gamma/\nu} F(\Delta^\nu L)
  \label{eq:27}
\end{equation}
we are led in finite systems to the presence of a peak in $ \chi(p, L)$,
located at $p_c(L)$ shifted from the infinite size critical point as
\begin{equation}
  |p_c(L) - p_c| \sim L^{-1/\nu}.
  \label{eq:5}
\end{equation}
The value of the susceptibility at this peak scales as 
\begin{equation}
  \chi(p_c(L)) \sim  L^{\gamma/\nu},
\end{equation}
while the order parameter scales as
\begin{equation}
  \mathcal{P}(p_c(L)) \simeq L^{-\beta/\nu}.
\end{equation}

In the case of networks, the application of this procedure is hindered
by the impossibility of defining a cluster to be spanning, and thus
distinguishing between percolating and finite clusters.  A different
approach is thus often followed~\cite{Colomer-de-Simon2014} based on the
fluctuations of the order parameter.  We analyze here this approach,
which can be applied to study also the SIR model, [while the one based
on $\chi$ (Eq.~(\ref{eq:16})) obviously cannot, because for SIR
only one cluster per run is generated].  Let us define the order parameter
\begin{equation}
  \phi = \frac{S}{N}
\end{equation}
where $S$ is the size of the largest cluster. In the limit $N\to\infty$,
for $p \leq p_c$, there is no giant component and $S$ is the size of a
finite component, so that $\phi \to 0$.  For $p>p_c$ instead, $S = G$,
and thus $\phi = \mathcal{P}$ is finite.

From this quantity and its moments, different definitions of
susceptibility, aiming at determining the critical point and associated
critical exponents, can be considered:
\begin{itemize}
\item Standard susceptibility in non-equilibrium phase transitions
  \cite{Marrobook}
  \begin{equation}
    \chi_1 = N [\av{\phi^2} - \av{\phi}^2] = \frac{\av{S^2}-\av{S}^2}{N} 
    \label{eq:21}
  \end{equation}

\item Susceptibility proposed for epidemic processes in networks
  \cite{Ferreira12,Colomer-de-Simon2014,PhysRevE.91.010801}
  \begin{equation}
    \chi_2 = N \frac{\av{\phi^2} - \av{\phi}^2}{\av{\phi}} = 
    \frac{\av{S^2} - \av{S}^2}{\av{S}}
    \label{eq:25}
  \end{equation}

\item Epidemic variability \cite{Crapey2006,Shu15}
  \begin{equation}
    \chi_3' = \frac{\sqrt{\av{\phi^2} - \av{\phi}^2}}{\av{\phi}} =
    \sqrt{\frac{\av{S^2}}{\av{S}^2} - 1} 
  \end{equation}
  Inspired by this definition, we will consider here the simplified form
  \begin{equation}
    \chi_3 = \frac{\av{\phi^2}}{\av{\phi}^2} =
    \frac{\av{S^2}}{\av{S}^2}.
    \label{eq:17}
  \end{equation}

\end{itemize}
In all previous definitions brackets $\langle \cdot \rangle$ indicate
averaging over different realizations of the percolation process.
We now analyze the suitability of each of these quantities as detectors
of criticality, by checking whether they fulfill the requirement
that they show a maximum close to the critical point,
whose position tends to $p_c$ while the height diverges as the system
size $N$ grows.

All three susceptibilities defined above ($\chi_1, \chi_2, \chi_3$) tend
to a finite value for all values of $p > p_c$ as the system size
diverges, because $\av{S} \sim N$, $\av{S^2} \sim N^2$ and the
fluctuations are Gaussian $\av{S^2}-\av{S}^2 \sim N$.  In the opposite
limit $p \to 0$, considering $p$ of the order of $N^{-1}$, we have that
all moments $\av{S^k}$ are independent of $N$ (since essentially the
number of edges added does not depend on $N$), so that $\chi_2$ and
$\chi_3$ go to a constant for $p\to0$, while $\chi_1$ goes to zero.

Let us now analyze the behavior at the critical point for large $N$.  At
criticality, the largest component in Euclidean lattices coincides with
the incipient spanning cluster $G_i$ \cite{stauffer94}, that is, the
spanning cluster observed at the percolation threshold, whose size
scales as a power of the system size,
$\av{G_i} \sim L^{d-\beta/\nu}$. In Ref.~\cite{Coniglio1980}, it is
proven that, in a regular $d$-dimensional lattice of size $L$
($N=L^d$) one has
\begin{equation}
  \frac{\av{G_i^2} -\av{G_i}^2}{N} \sim \frac{\av{G_i^2}}{N} \sim
  \frac{\av{G_i}^2}{N} \sim L^{\gamma / \nu},  \;  \frac{\av{G_i}}{N} \sim
  L^{-\beta / \nu}.
\end{equation}
Assuming that the same scaling laws can be extended to the behavior of
the largest cluster size $S_c$ at the percolation threshold in networks,
with the system size $L$ replaced by the network size $N$, we have
\begin{equation}
  \frac{\av{S_c^2} -\av{S_c}^2}{N} \sim \frac{\av{S_c^2}}{N} \sim
  \frac{\av{S_c}^2}{N} \sim N^{\gamma / \nu}, \quad  \frac{\av{S_c}}{N} \sim
  N^{-\beta / \nu}, 
  \label{eq:3}
\end{equation}
with the corresponding change in the definition of the exponent
$\nu$. As we will see below, the previous scaling forms are confirmed by 
percolation simulations in random networks.  The scaling relations
in Eq.~(\ref{eq:3}) can be also obtained by assuming the so-called first
scaling law~\cite{1742-6596-297-1-012005} for the probability
distribution of the order parameter at criticality
\begin{equation}
 P(S_c) = \frac{1}{\av{S_c}}F(S_c/\av{S_c}),
  \label{eq:6}
\end{equation}
which implies
\begin{equation}
  \av{S_c^k} \sim \av{S_c}^k.
\label{eq:7}
\end{equation}
This leads to the results in Eq.~(\ref{eq:3}), assuming
$\av{S_c}/N \simeq N^{- \beta/\nu}$ and the hyperscaling relation
\begin{equation}
  \frac{2 \beta}{\nu} + \frac{\gamma}{\nu} = 1.
  \label{eq:20}
\end{equation}

Inserting Eqs.~(\ref{eq:3}) into the definitions of the susceptibilities
we obtain the behavior at criticality
\begin{equation}
  \chi_1(p_c) \sim  N^{\gamma/\nu },\quad
  \chi_2(p_c)  \sim  N^{(\gamma + \beta)/\nu}, \quad
  \chi_3(p_c) \sim \mathrm{const}.  
  \label{eq:22}
\end{equation}
The previous relationships show that $\chi_1$ and $\chi_2$ are suitable
criticality detectors: They diverge at the critical point, while tending
to a constant value away from criticality.  In this respect, $\chi_2$
should be numerically preferred, as it diverges with a larger exponent.
The susceptibility $\chi_3$ instead does not depend on $N$ at
criticality.  This makes it rather unsuitable as criticality detector in
the usual sense of a susceptibility with a diverging peak; however
$\chi_3$ might play in this case a role analogue to
Binder's cumulant for Ising-like equilibrium
transitions~\cite{Binder81,Binderbook,Dickman98}: a function of the
control parameter $p$ changing with the system size $N$ for all values
of $p$ except $p_c$, so that the latter is the estimated as the value
where curves of $\chi_3(p)$, computed for different of $N$, intersect
each other.

We have checked the performance of these three different
criticality detectors by performing bond percolation experiments using the
Newman-Ziff algorithm \cite{Newman2000,NewmanZiff2001} on two examples
of networks for which exact values of the percolation point and critical
exponents are available: Random regular networks and scale-free networks
generated with the uncorrelated configuration model (UCM)
\cite{Catanzaro05}. In random regular networks (RRN) all nodes have the
same degree $K$,
with edges randomly distributed among them, preventing self-connections
and multiple connections. The critical point is, according to
Eq.~(\ref{eq:1}),
\begin{equation}
  \label{eq:RRNPercoCritical}
  p_c = \frac{1}{K-1},
\end{equation}
while the values of the associated critical exponents are, from
Eqs.~(\ref{eq:9}), (\ref{eq:19}), and~(\ref{eq:220}),
$\beta_\mathrm{th}=\gamma_\mathrm{th}=1$, and $\nu_\mathrm{th}=3$.  In
our simulations, we fix $K=5$, leading to the theoretical critical point
$p_c^\mathrm{RRN, th} = 0.25$.  For scale-free networks, we consider a
degree exponent $\gamma_d = 3.5$, with a minimum degree
$k_\mathrm{min}=3$ and a maximum degree
$k_\mathrm{max} = N^{1/{(\gamma_d-1)}}$, equal to the so-called natural
cut-off~\cite{mariancutofss}. In
this case, the percolation threshold takes the form of
Eq.~(\ref{eq:1}).  Considering a pure discrete power-law form
$P(k) = k^{-\gamma} / \sum_{q=k_\mathrm{min}}^\infty q^{-\gamma}$, we
obtain $p_c^\mathrm{UCM,th} = 0.15054$. For this degree exponent, from
Eqs.~(\ref{eq:9}), (\ref{eq:19}), and~(\ref{eq:220}), we have
$\beta_\mathrm{th} = 2$, $\gamma_\mathrm{th} = 1$, and
$\nu_\mathrm{th} = 5$. 
In our simulations, the moments of the largest
cluster $\av{S^k}$ are computed averaging over $10000$ bond percolation
realizations on fixed networks of different size.

\begin{figure}[t]
  \centering
  \includegraphics[width=7cm]{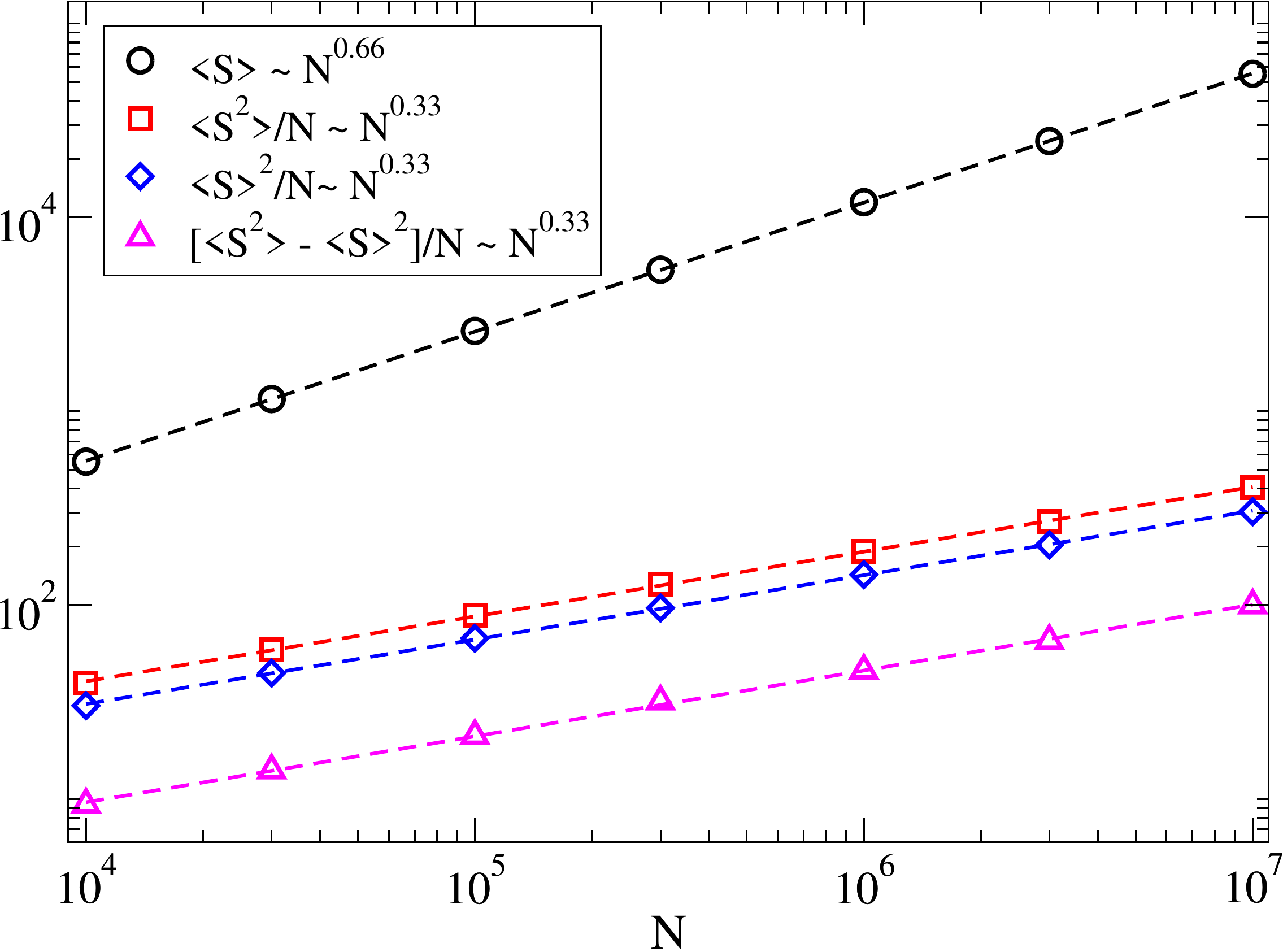}
  \caption{Numerical check of the scaling relations in Eq.~(\ref{eq:3})
    on RRN with fixed degree $K=5$.}
  \label{scalings}
\end{figure}

In the first place, we proceed to verify that the scaling relations in
Eqs.~(\ref{eq:3}) are observed numerically for percolation on RRN
networks. Thus, in Fig.~\ref{scalings} we plot different moments of the
distribution of the largest cluster size, computed at the 
theoretical critical point
$p_c^\mathrm{RRN,th} = 0.25$.
As we can see, the scaling relations assumed in Eq.~(\ref{eq:3}) are
perfectly satisfied, within the numerical accuracy of our
simulations.

\begin{figure}[t]
\centering
\includegraphics[width=\columnwidth]{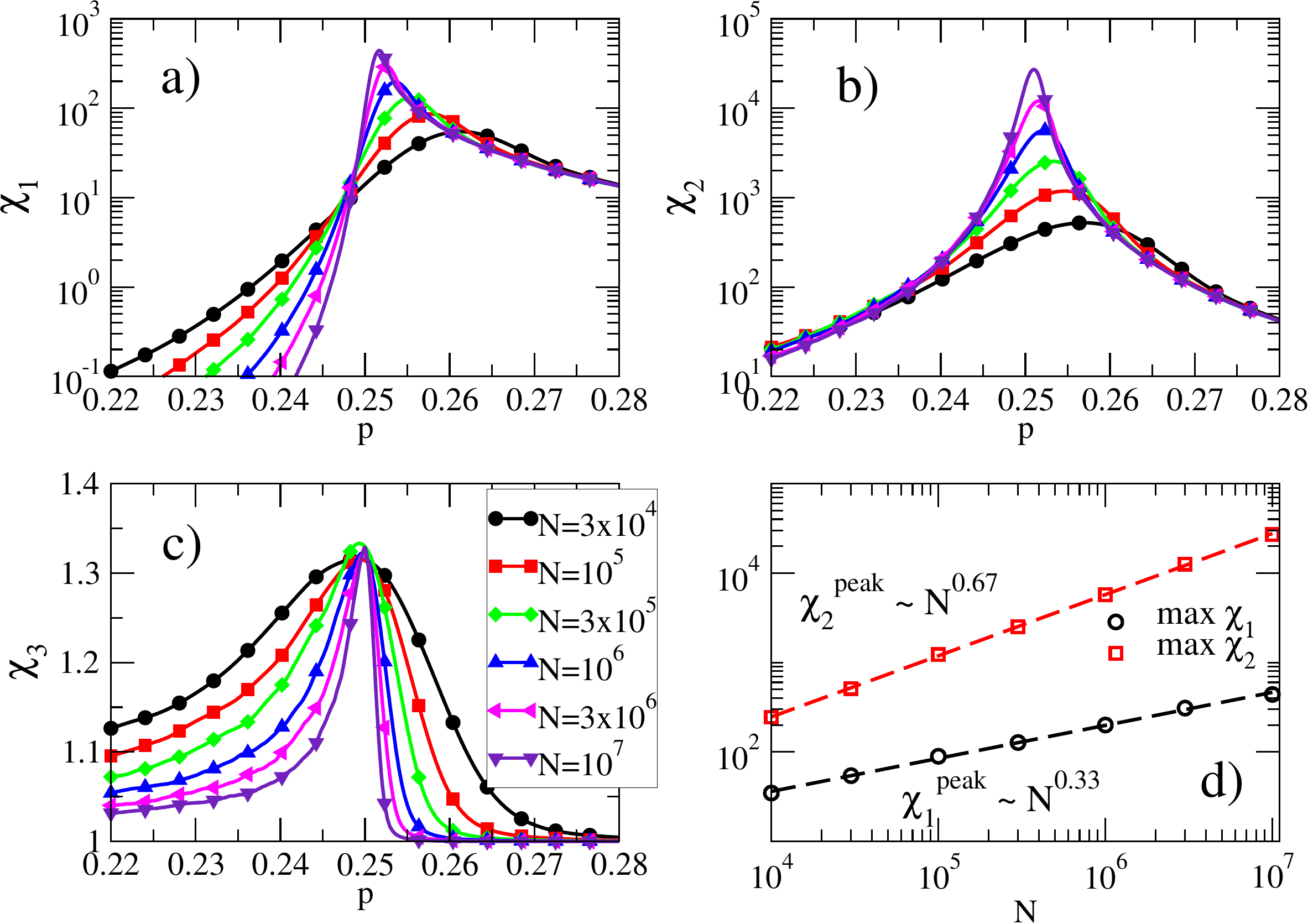}
\caption{Panels (a,b,c):
Different susceptibilities for bond percolation in RRN networks
  with degree $K=5$. Panel (d): Scaling of the height of peaks
  of the susceptibilities $\chi_1$ and $\chi_2$ as a function of the
  network size.}
\label{fig:2}
\end{figure}

\begin{figure}[t]
\centering
\includegraphics[width=\columnwidth]{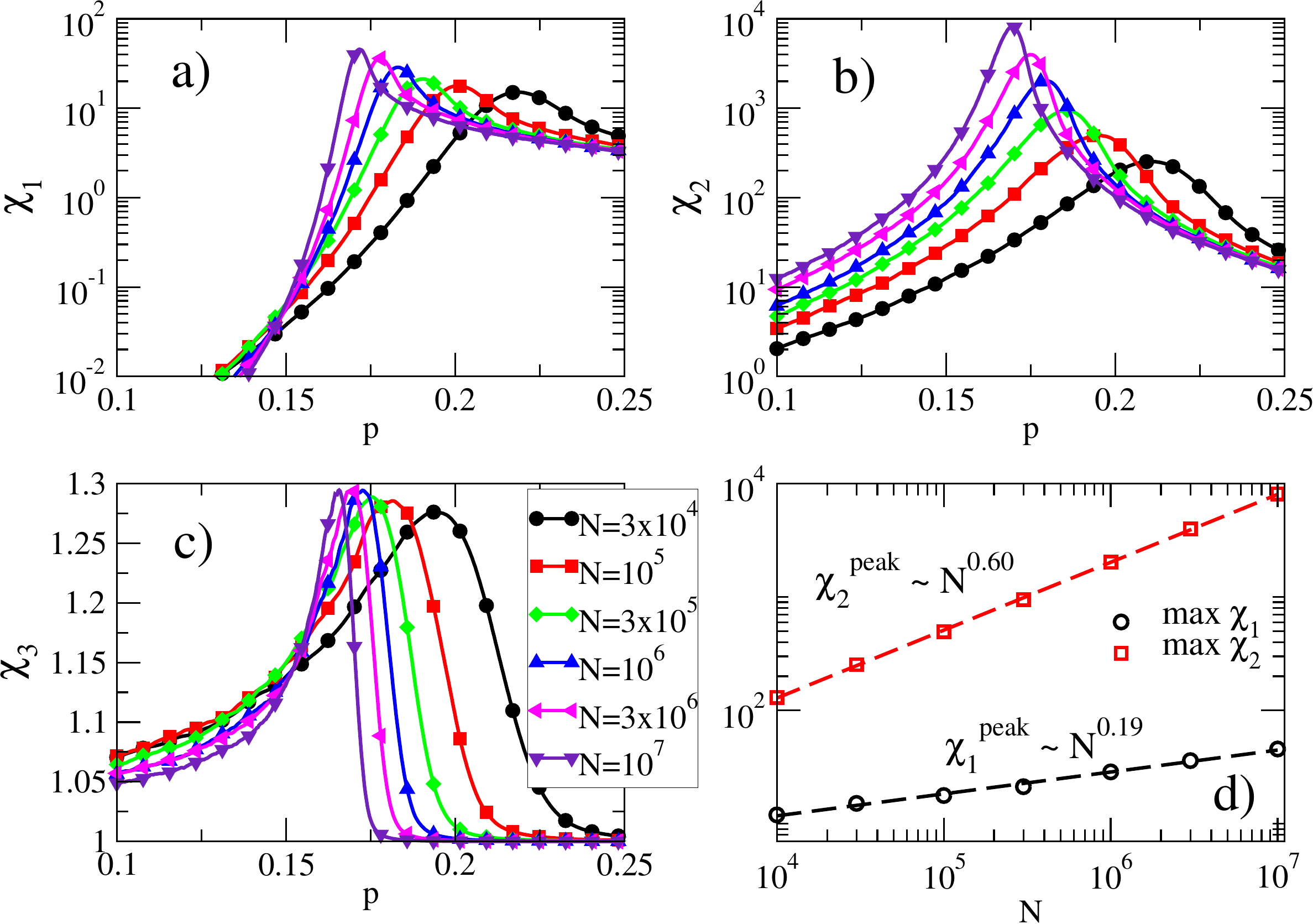}
\caption{Panels (a,b,c):
  Different susceptibilities for bond percolation in UCM networks
  with degree exponent $\gamma_d = 3.5$. Panel (d): Scaling of the
  height of peaks of the susceptibilities $\chi_1$ and $\chi_2$ as a
  function of the network size.}
\label{fig:2bis}
\end{figure}

We next plot the different susceptibilites as a function of $p$ for
different network sizes in the case of RRN, Fig.~\ref{fig:2}, and UCM
networks, Fig.~\ref{fig:2bis}. As we can see, in both cases $\chi_1$ and
$\chi_2$ show peaks of height increasing with $N$, located at positions
$p_c(N)$ that change with network size, moving with increasing $N$
towards smaller $p$ values.
From the divergence of the height of the peaks of the susceptibilities
we can obtain the values of exponent ratios involving $\gamma$.  Indeed,
assuming that the susceptibilities $\chi_1$ and $\chi_2$ obey the FSS
form (see Eq.~(\ref{eq:27}))
\begin{equation}
  N^{- \alpha_i / \nu}  \chi_i(p, N) =  F_i[(p - p_c)N^{1/\nu}],
  \label{eq:28}
\end{equation}
where $\alpha_1 = \gamma$ and $\alpha_2 = \gamma + \beta$, we obtain
that the height of the susceptibilities at their peak,
$\chi_i^\mathrm{peak} = \chi_i(p_c(N))$, must satisfy
\begin{equation}
  \chi_1^\mathrm{peak} \sim N^{\gamma/\nu}, \quad  
  \chi_2^\mathrm{peak} \sim N^{(\gamma+\beta)/\nu}.
\label{eq:26}
\end{equation}
From a linear regression in logarithmic scale of the peak height as a
function of $N$, we obtain for RRN (Fig.~\ref{fig:2}(d)) the exponent
ratios $\gamma / \nu = 0.33(1)$, $(\beta + \gamma)/\nu = 0.67(1)$, which
compare very well with the theoretical values
$\gamma_\mathrm{th} / \nu_\mathrm{th} = 1/3$ and
$(\gamma_\mathrm{th} + \beta_\mathrm{th}) / \nu_\mathrm{th} = 2/3$.  For
UCM networks (Fig.~\ref{fig:2bis}(d)) we find $\gamma/\nu = 0.19(1)$,
$(\gamma+\beta)/\nu = 0.60(1)$, in excellent agreement with the
theoretical expectations $\gamma_\mathrm{th}/\nu_\mathrm{th} = 0.2$ and
$(\gamma_\mathrm{th}+\beta_\mathrm{th})/\nu_\mathrm{th} = 0.6$.

\begin{figure}[t]
\centering
\includegraphics[width=7cm]{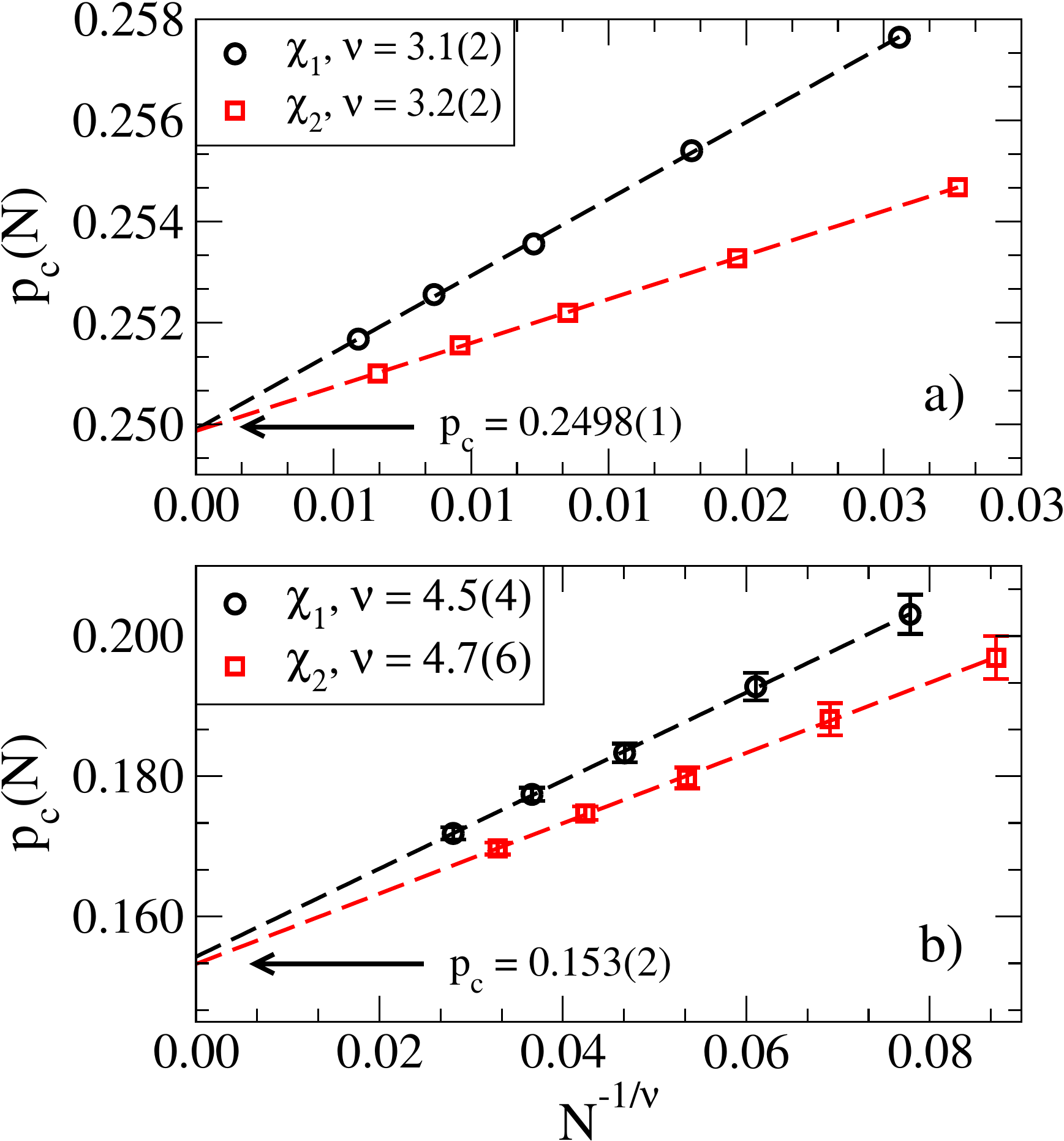}
\caption{Plot of the peak position as a function of $N$ for bond
  percolation on RRN (a) and UCM networks with $\gamma_d=3.5$ (b). The
  linear behavior is in agreement with Eq.~(\ref{eq:31}).}
\label{fig:peaks}
\end{figure}

From the positions of the peaks $p_c(N)$ as a function of $N$ we can
obtain information on the asymptotic critical point (in the infinite
network size limit) and the exponent $\nu$, assuming the validity of
Eq.~(\ref{eq:5}). In this case, we can write
\begin{equation}
  p_c(N) = p_c - a N^{-1/\nu},
  \label{eq:31}
\end{equation}
where $a$ is some constant prefactor. By means of a non-linear fitting
of data to Eq.~(\ref{eq:31}), the values of $p_c$ and $\nu$ can be
estimated. From such a non-linear fitting, we obtain for RRN the value
$\nu = 3.1(2)$ for $\chi_1$ and $\nu=3.2(2)$ for $\chi_2$, with a
critical point $p_c = 0.2498(1)$ coincident for both susceptibilities,
see Fig.~\ref{fig:peaks}.
In the case of RRN networks, a single network sample is sufficient, due
to the fact that the position of the peaks $p_c(N)$ fluctuates very
slightly from sample to sample. These fluctuations are stronger in UCM
networks, so we proceed to estimate the peak in 10 different samples of
networks of given size $N$, and compute from them the average position
$p_c(N)$ and associated error, see Fig.~\ref{fig:peaks}. Applying to
this data a non-linear fitting to the form of Eq.~(\ref{eq:31}), we
obtain $\nu = 4.5(4)$ for $\chi_1$ and $\nu=4.7(6)$ for $\chi_2$, with a
common critical point $p_c=0.153(2)$. The values thus obtained show a
very good match with the theoretical expectations for the RRN,
$p_c^\mathrm{RRN, th} = 0.25$ and $\nu_\mathrm{th} = 3$, and provide a
quite reasonable approximation in the case of UCM networks,
$p_c^\mathrm{RRN, th} = 0.15054$ and $\nu_\mathrm{th} = 5$.

In order to check the accuracy of the different susceptibilities with
respect to the known exact values of the critical point and critical
exponents, we perform a data collapse analysis. The validity of FSS
hypothesis above implies that plotting the values of the
susceptibilities rescaled according to Eq.~(\ref{eq:28}) curves for
different values of $N$ will collapse onto the same universal function
$F_i(x)$ for $x=(p-p_c) N^{1/\nu}$, when the correct values of the
critical points and critical exponents are used. 
\begin{figure}[t]
  \centering
  \includegraphics[width=7cm]{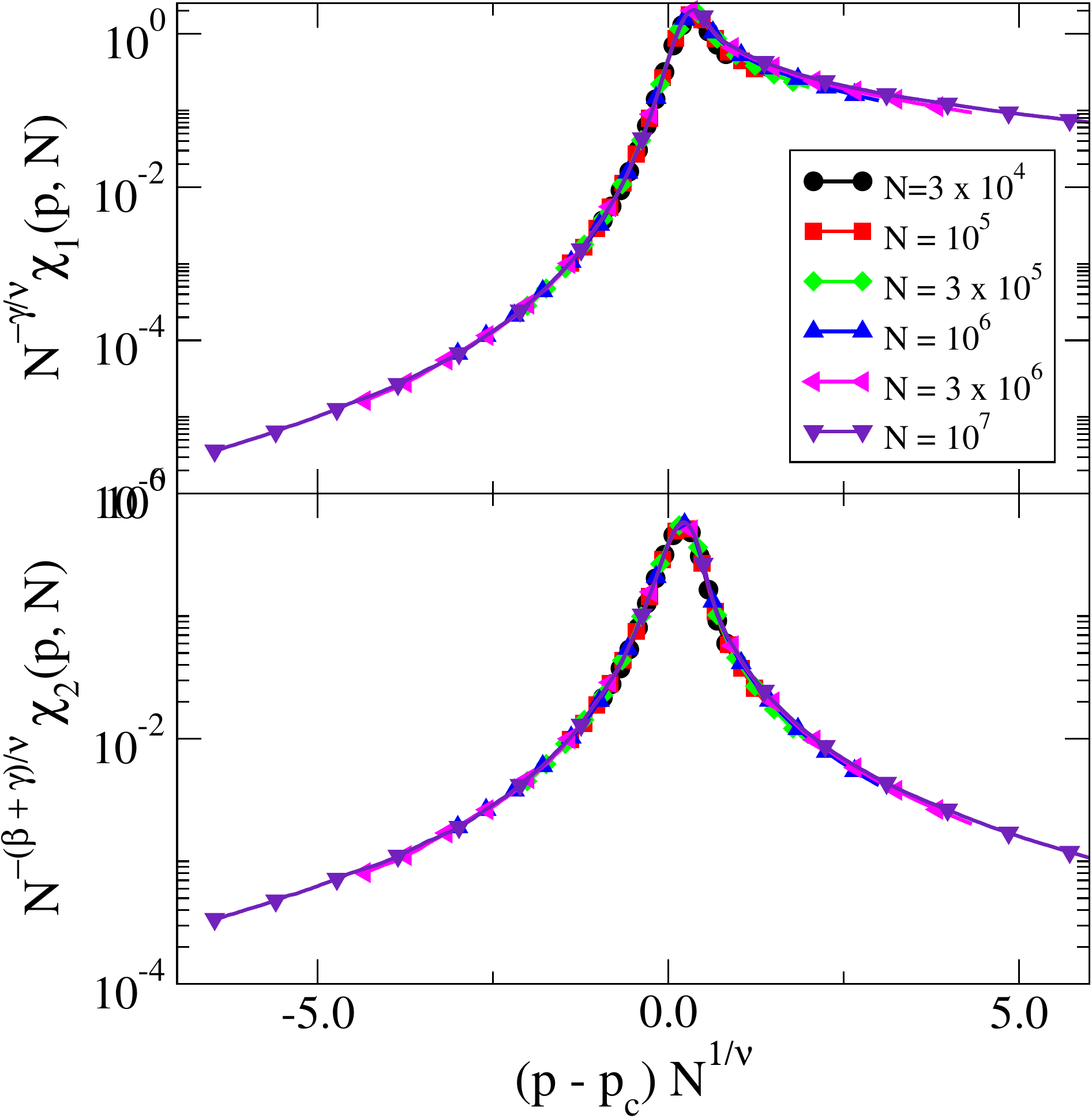}
  \caption{Data collapse analysis of the susceptibilities $\chi_1$ (top)
    and $\chi_2$ (bottom) for bond percolation on RRN of degree
    $K=5$. We have used the exact theoretical values
    $p_c^\mathrm{RRN, th} = 0.25$, $\nu_\mathrm{th}=3$,
    $\gamma_\mathrm{th} / \nu_\mathrm{th} = 1/3$,
    $(\beta_\mathrm{th} + \gamma_\mathrm{th}) / \nu_\mathrm{th} = 2/3$.}
  \label{fig:collapseRRN}
\end{figure}
In Fig.~\ref{fig:collapseRRN} we show the data collapse analysis for the
RRN. In this case, a perfect data collapse is obtained with the exact
theoretical results $p_c^\mathrm{RRN,th} = 0.25$, $\nu_\mathrm{th}=3$,
$\gamma_\mathrm{th} / \nu_\mathrm{th} = 1/3$,
$(\beta_\mathrm{th} + \gamma_\mathrm{th}) / \nu_\mathrm{th} = 2/3$.
\begin{figure}[t]
  \centering
  \includegraphics[width=7cm]{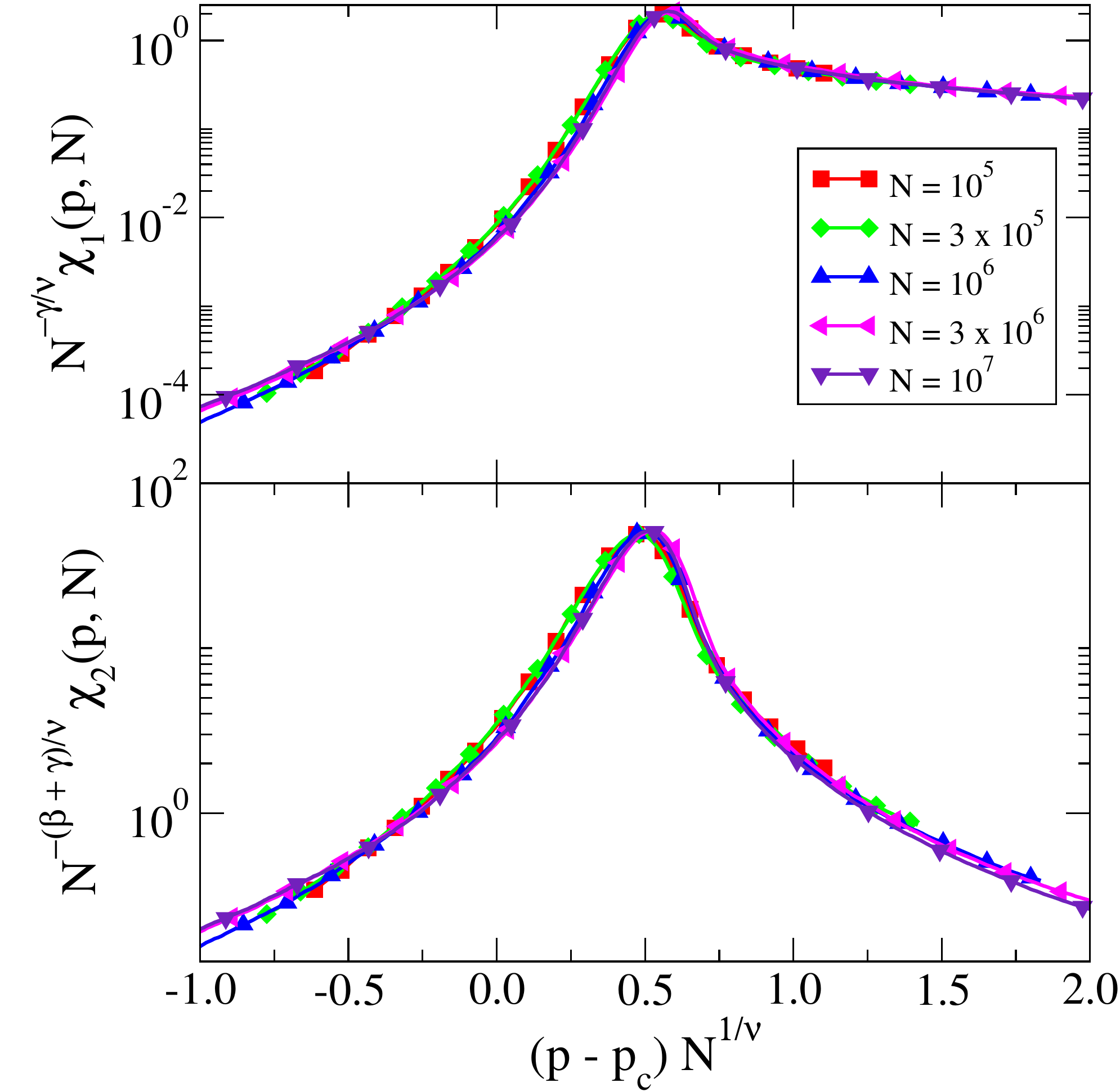}
  \caption{Data collapse analysis of the susceptibilities $\chi_1$ (top)
    and $\chi_2$ (bottom) for bond percolation on UCM of degree exponent
    $\gamma_d = 3.5$. We have used the numerical critical point
    $p_c^\mathrm{UCM} = 0.153$, and the exponents $\nu = 4.7$,
    $\gamma/\nu = 0.19$ and $(\gamma+\beta)/\nu = 0.60$.}
  \label{fig:collapseUCM}
\end{figure}
Concerning the scale-free UCM networks, a very good data collapse is
obtained using the numerical parameters previously estimated from the
analysis of the peak height and peak position of the susceptibilities,
namely $p_c^\mathrm{UCM} = 0.153$, $\gamma/\nu = 0.19$,
$(\gamma+\beta)/\nu = 0.60$, and $\nu = 4.7$.

\begin{figure}[t]
  \centering
  \includegraphics[width=\columnwidth]{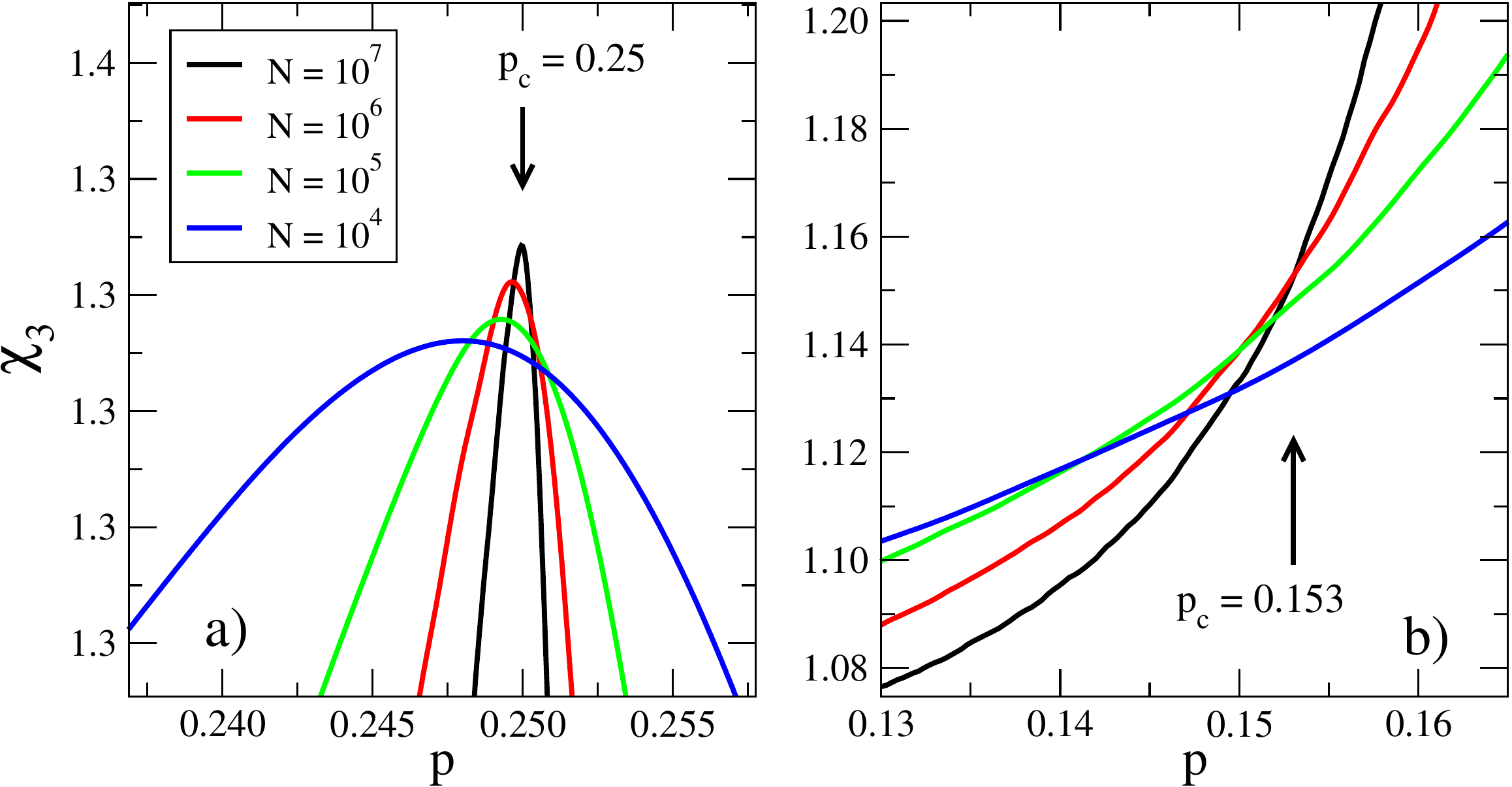}
  \caption{Close up of the susceptibility $\chi_3$ in the vicinity of
    the critical point for RRN networks (a) and UCM networks with
    $\gamma_d=3.5$ (b). In this last case, we consider the average over
    $10$ different network samples.}
  \label{crossingzooms}
\end{figure}

As we have pointed out above, one could think on using the constant
value of the susceptibility $\chi_3$ as a method to determine the
critical point in the sense of the Binder cumulant: Since $\chi_3(p_c)$
does not depend on network size, curves of $\chi_3(p)$ for different
values of $N$ should cross exactly at $p_c$, allowing thus to identify
$p_c$. The usefulness of this method, however, is hindered by the fact
that, at odds with the originally defined Binder cumulant, $\chi_3(p)$
has in general an asymptotic form that is not a step function
\cite{Binder81}: The limits for large and small values of $p$ are very
similar, and the function exhibits a peak close to $p_c$. In the case of
RRN, see a close up in the vicinity of the critical point in
Fig.~\ref{crossingzooms}(a), the peaks are so close to the critical
point that in general two intersection points can be observed for every
pair of curves, rendering them unsuitable for the determination of
$p_c$. In the case of UCM networks with $\gamma_d = 3.5$,
Fig.~\ref{fig:2bis}(b), the crossing is sufficiently away from the peak
to allow an estimate of the crossing point which is in reasonable
agreement with the estimated numerical one,
$p_c^\mathrm{UCM} \simeq 0.153$, for the largest network sizes
considered, see the corresponding close up in
Fig.~\ref{crossingzooms}(b).

\subsection{SIR model}
\label{sec:sir-model-analysis}

The SIR process is mapped exactly to bond percolation. However, when the
two processes are simulated numerically, there is a crucial difference:
In a percolation experiment, we have information on the whole cluster
structure for each percolation configuration, and we can pick the
largest cluster to perform statistics. On the other hand, in the SIR
process, we obtain only one outbreak, corresponding to a particular
percolation cluster, in every run. Therefore, averages over different 
outbreak sizes are the only available information.
Moreover, since the seed of SIR outbreaks is chosen randomly among all 
vertices,
epidemic outbreaks are bond percolation clusters {\em sampled with a
  probability proportional to their size}.  In other words, when an
outbreak occurs above the critical point, this corresponds to the giant
component, of size $G$, with probability $P_G = G/N$, and to a finite
cluster with probability $P_F = 1- P_G$. On the other hand, when the
system is below the critical point, all outbreaks correspond to finite
clusters. Thus, if we define the order parameter as the relative
outbreak size, $\phi = N_R / N$, when computing its moments in SIR
simulations we are performing, in the general case, an implicit double
average:
\begin{enumerate}
\item For a fixed percolation configuration, we are selecting the giant
  component with probability $P_G$, and finite clusters with probability
  $1-P_G$\footnote{We assume $P_G = G = 0$ below the critical
    point.}. In the latter case, a finite cluster of size $s$ is
  selected with probability $\sim s n_s$.  Importantly, this first
  average is made at constant $G$ (the size of the giant component of
  the fixed percolation configuration).
  
  For the fixed percolation configuration, the $n$-th moment of the
  order parameter $\phi$ is thus
    \begin{eqnarray}
      \overline{\phi^n}& =& \left( \frac{G}{N}\right)^{n} P_G +
                            \sum_{s<G} \left( 
        \frac{s}{N}\right )^n \frac{s n_s}{\sum_{s' < G} s' n_{s'}} P_F
                            \nonumber
      \\
      &=&
      \left( \frac{G}{N}\right)^{n+1} + \mu \sum_{s<G} \left(
        \frac{s}{N}\right)^n s n_s \left ( 1 - 
        \frac{G}{N} \right)
    \end{eqnarray}
  where $\mu^{-1} = \sum_{s' < G} s' n_{s'}$ is a normalization factor
  that tends to a constant in the limit of large $N$.

\item 
  After this average, an average over different
  percolation configurations, described by the distribution of giant
  component sizes $g(G)$, must be performed. Thus we have
    \begin{eqnarray*}
      \av{  \overline{\phi^n}} &=& \av{ \left( \frac{G}{N}\right)^{n+1}
                                   }_g + \mu 
      \sum_{s<G} \left( \frac{1}{N}\right)^n \av{ s^{n+1} n_s \left ( 1 -
          \frac{G}{N} \right)}_g \\
      &\equiv& \sum_S  \left(
        \frac{G}{N}\right)^{n+1} g(G) + \mu \sum_{s<G} \left( \frac{1}{N}\right)^n \av{ s^{n+1} n_s \left ( 1 -
          \frac{G}{N} \right)}_g
    \end{eqnarray*}

\end{enumerate}
Assuming that the largest cluster and the finite clusters are
uncorrelated ($\langle s^{n+1} G \rangle = \langle s^{n+1}\rangle 
\langle G\rangle$) 
we have
\begin{equation}
  \av{\overline{\phi^n}}  = \frac{\av{G^{n+1}}}{N^{n+1}} + \mu
  \frac{\av{s^{n+1}}}{N^n} \left( 1- \frac{\av{G}}{N}\right)
\label{final}
\end{equation}
where the averages $\av{s^k}$ of finite clusters are performed with the
probability $n_s(p)$.

Let us analyze the scaling of the three candidate susceptibilities at
the critical point.  Considering first the approach to criticality from
below. Since there is no giant component, we have
\begin{equation}
  \av{  \overline{\phi^n}}  = 
  \frac{\av{s^{n+1}}}{N^n} .
\end{equation}
Below the critical point, the moments $\av{s^k}$ can be computed from
the scaling ansatz for the normalized cluster number $n_s(p)$, see
Eqs.~(\ref{eq:4}) and~(\ref{betagamma}), leading to
\begin{equation}
  \av{s^2} \simeq \Delta^{-\gamma}, \quad
  \av{s^3} \simeq \Delta^{-(4-\tau)/\sigma} \simeq \Delta^{-(\beta + 2\gamma)}.
\end{equation}
Applying the FSS hypothesis, substituting $\Delta \sim N^{-1/\nu}$, we
have, at criticality,
\begin{equation}
  \av{s^2}_c \simeq N^{\gamma/\nu}, \quad  \av{s^3}_c \simeq N^{(\beta +
    2\gamma)/\nu} ,
\end{equation}
and from here
\begin{equation}
    \av{  \overline{\phi}}_c \simeq  N^{\gamma/\nu -1}, \quad 
    \av{  \overline{\phi^2}}_c \simeq  N^{(\beta + 2\gamma)/\nu -2}.
\label{critscal}
\end{equation}
Therefore, we have
\begin{equation}
  \chi_1(\lambda_c) = N [\av{\overline{\phi^2}}_c - \av{\overline{\phi}}^2_c] \simeq
  N^{(\beta +  2\gamma)/\nu -1} \simeq  N^{(\gamma-\beta)/\nu},
\end{equation}
where the hyperscaling relation Eq.~(\ref{eq:20}) has been used.
For $\chi_2(\lambda_c)$, in the limit of large $N$, we have
\begin{equation}
  \chi_2(\lambda_c) = \frac{\chi_1(\lambda_c)}{\av{\overline{\phi}}_c} \simeq N^{1-\beta/\nu},
\end{equation}
and, finally, for $\chi_3$
\begin{equation}
  \chi_3(\lambda_c) = \frac{\av{\overline{\phi^2}}_c}{\av{\overline{\phi}^2})c} \simeq
  N^{\beta/\nu} .
\end{equation}

Results for the same quantities can be derived when approaching 
criticality from above.  In this case, $\av{G^k}$ has the leading
behavior and terms $\av{s^k}$ can be disregarded. The
definitions of the candidate susceptibilities become therefore
\begin{equation}
  \chi_1 = N [\av{\overline{\phi^2}} - \av{\overline{\phi}}^2] = 
  \frac{\av{G^3}-\av{G^2}^2/N}{N^2}
\label{chi1sir}
\end{equation}
\begin{equation}
  \chi_2 = N \frac{\av{\overline{\phi^2}} -
    \av{\overline{\phi}}^2}{\av{\overline{\phi}}} =  
  \frac{\av{G^3} - \av{G^2}^2/N}{\av{G^2}}
\end{equation}
\begin{equation}
  \chi_3 = \frac{\av{\overline{\phi^2}}}{\av{\overline{\phi}}^2} = 
\frac{N \av{G^3}}{\av{G^2}^2}
\end{equation}

For fixed $\lambda > \lambda_c$ and large $N$, the numerator of $\chi_1$
and $\chi_2$ increases as $N^3$ so that both $\chi_1$ and $\chi_2$ grow
linearly with $N$.  This is already enough to declare the two quantities
unsuitable as detectors of criticality. Instead it is trivial to see
that $\chi_3$ goes to a finite limit as $N \to \infty$.  Let us also
check the behavior at criticality.  Right at the critical point we
assume that the size of the largest component obeys
$\av{S_c^3} \sim \av{S_c^2} \av{S_c}$, which follows from the scaling
relation Eq.~(\ref{eq:6}). Hence, from Eq.~(\ref{eq:3}),
$\av{S_c^3} \sim N^{2+(\gamma-\beta)/\nu}$ so that
\begin{equation}
  \chi_1(\lambda_c) \simeq \frac{\av{S_c^3}}{N^2} \sim N^{(\gamma-\beta)/\nu}
\end{equation}
We conclude that this function is in general not a good detector of
criticality, since in general $\gamma \le \beta$, and therefore
$\chi_1(\lambda_c)$ decreases with network size.  It may however be of
use in the case $\gamma=\beta$ (as in MF), because in this case all
curves for different $N$ cross each other at the critical point, thus
allowing its identification.

\begin{figure}
  \centering
  \includegraphics[width=\columnwidth]{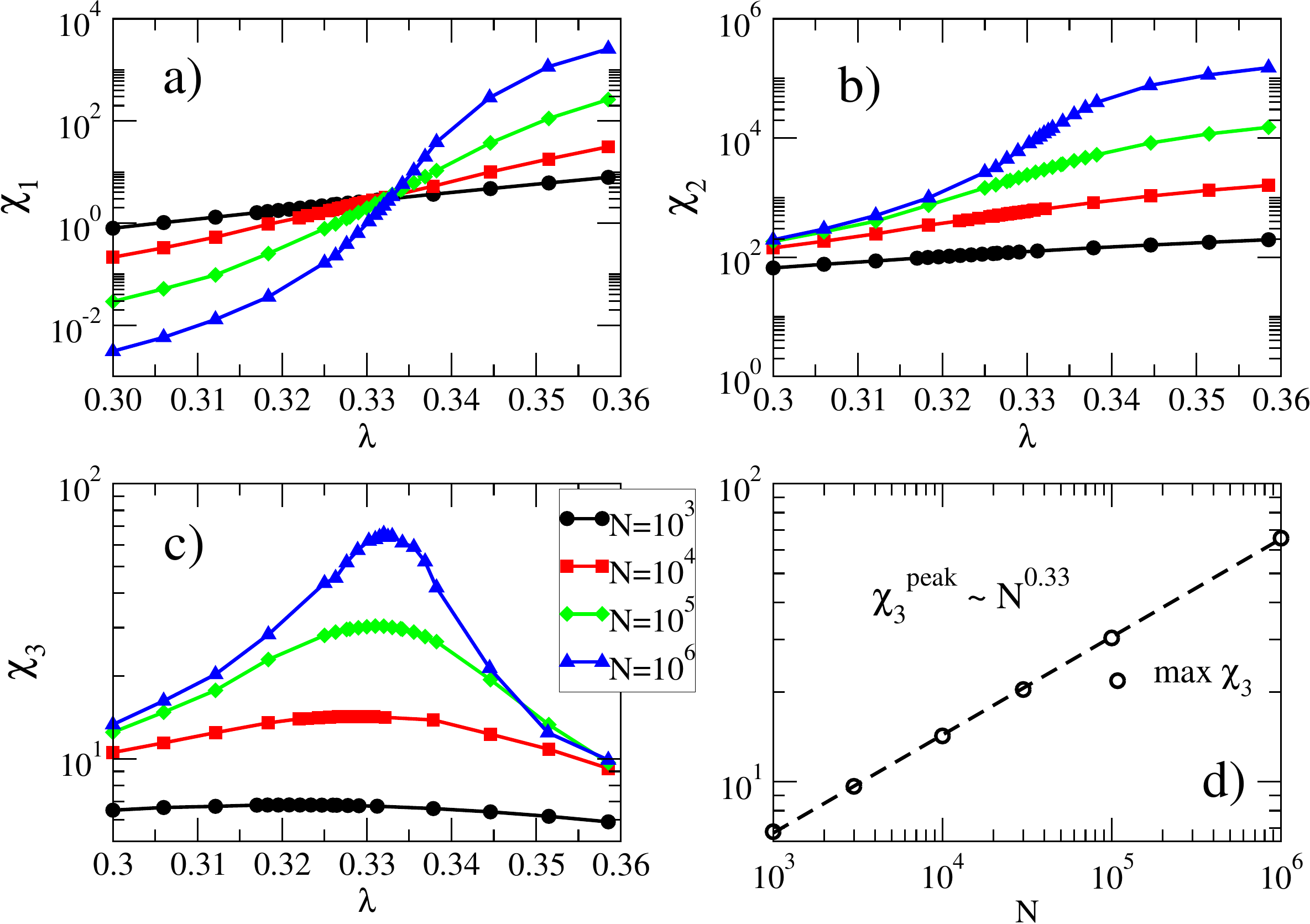}
  \caption{Panels (a,b,c): Different susceptibilities for the SIR
    process in RRN networks with degree $K=5$. Panel (d): Scaling of the
    height of peak of the susceptibility $\chi_3$ as a function of the
    network size.}
\label{fig:3}
\end{figure}

With regard to $\chi_2$, making the same assumptions about 
the scaling of $\av{S_c^3}$ and the irrelevance of $\av{S_c^2}^2/N$ 
we find now
\begin{equation}
  \chi_{2}(\lambda_c) \simeq \av{S_c} \sim N^{1-\beta/\nu}.
\end{equation}
Hence the value of $\chi_2(\lambda_c)$ grows at the critical point but, since
it grows even more strongly above the critical point, $\chi_2$ has no
maximum at $\lambda_c$ (see Fig.~\ref{fig:3}).  It is hence unsuitable
as detector of criticality.

Under the same assumptions about the behavior at criticality, we also
obtain
\begin{equation}
  \chi_{3}(\lambda_c) \simeq \frac{N \av{S_c}}{\av{S_c^2}} \sim N^{\beta/\nu}.
  \label{eq:13}
\end{equation}
It is therefore a good detector of criticality.

We have tested these predictions for the SIR model on RRN with fixed
degree $K=5$ and UCM scale-free networks with degree exponent
$\gamma_d = 3.5$. For RRN networks, the mean-field theoretical
prediction for the epidemic threshold is
$\lambda_c^\mathrm{RRN,th}=1/(K-2)=1/3$; in the case of UCM networks,
Eq.~(\ref{eq:12}), with a discrete power-law distribution
$P(k) = k^{-\gamma} / \sum_{q=k_\mathrm{min}}^\infty q^{-\gamma}$, leads
to the threshold $\lambda_c^\mathrm{UCM,th} = 0.1772$. In both cases,
the theoretical predictions for the critical exponents should be the
same as in percolation, namely $\beta_\mathrm{th}=\gamma_\mathrm{th}=1$,
and $\nu_\mathrm{th}=3$ for RRN networks, and $\beta_\mathrm{th} = 2$,
$\gamma_\mathrm{th} = 1$, and $\nu_\mathrm{th} = 5$ for UCM networks
with $\gamma_d = 3.5$.  In our simulations, the moments of the relative
outbreak size $\av{\phi^k}$ are computed averaging over at least $10000$
realizations (up to $10^7$ realizations close to the critical point) of
the epidemic process on fixed networks of different size.

\begin{figure}
  \centering
  \includegraphics[width=\columnwidth]{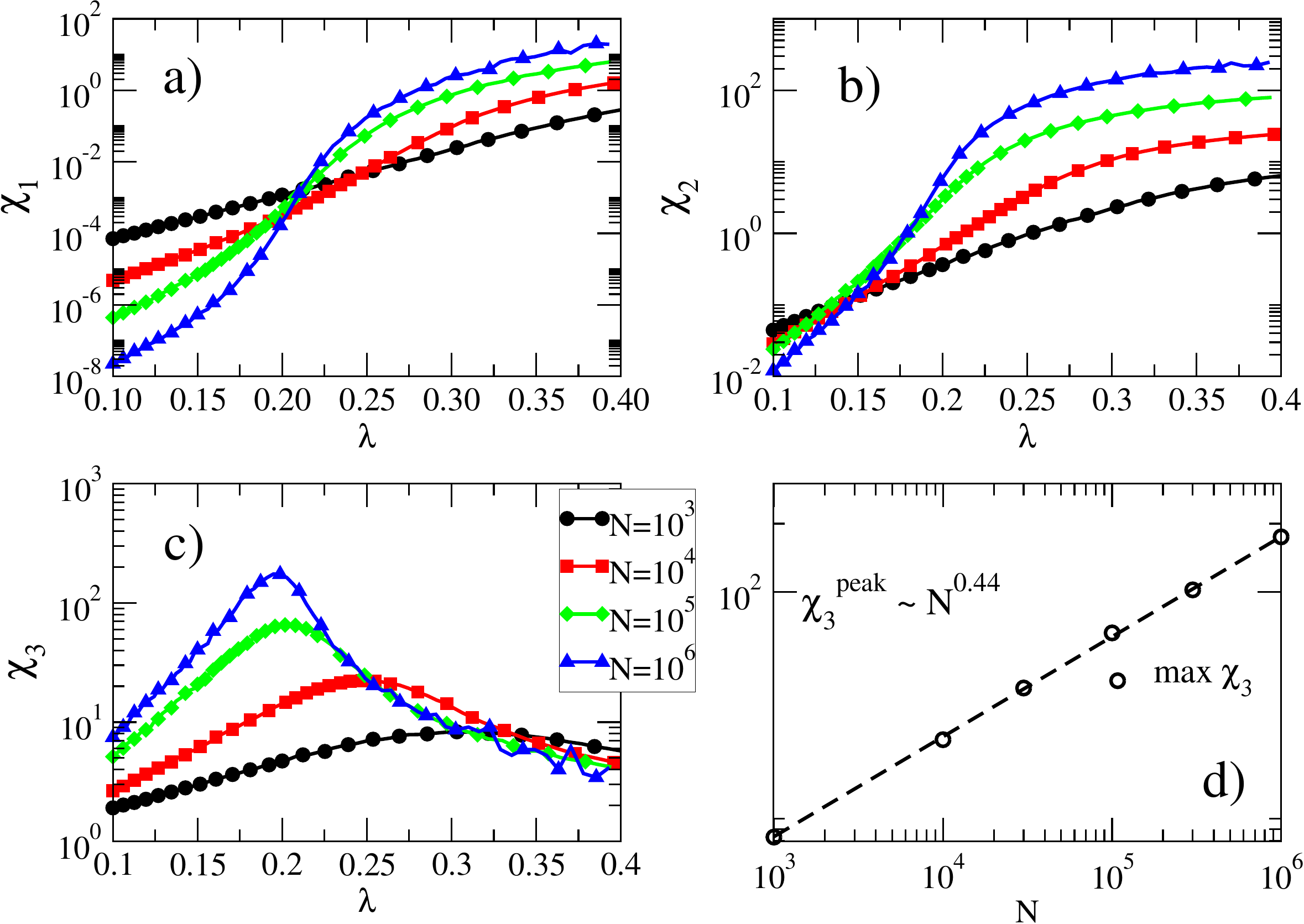}
  \caption{Panels (a,b,c): Different susceptibilities for the SIR
    process in UCM networks with degree exponent $\gamma_d = 3.5$. Panel
    (d): Scaling of the height of peak of the susceptibility $\chi_3$ as
    a function of the network size.}
  \label{fig:4}
\end{figure}

In Figs.~\ref{fig:3} and~\ref{fig:4} we plot the three
susceptibilities as a function of the spreading rate $\lambda$ in
different network sizes, for the RRN and UCM cases,
respectively. Figures shows that $\chi_3$ exhibits a well pronounced
maximum, $\lambda_c(N)$, growing with $N$; $\chi_2$ instead does not
possess a maximum. In the case of RRN, where $\beta = \gamma$, $\chi_1$
does not exhibit a maximum either, but the critical point can be
detected as the point where curves for different values of $N$
meet. This is not possible in UCM networks, where $\beta > \gamma$.

Assuming a FSS hypothesis for $\chi_3$ of the form
\begin{equation}
  N^{-\beta/\nu} \chi_3(\lambda, N) = F[(\lambda-\lambda_c)N^{1/\nu}],
  \label{eq:2}
\end{equation}
implies that the height of the susceptibility peak
$\chi_3^\mathrm{peak} = \chi_3(\lambda_c(N))$, scales as
\begin{equation}
  \chi_3^\mathrm{peak} \sim N^{\beta/\nu}.
\end{equation}
From here, using a linear regression in logarithmic scale, we obtain the
estimates $\beta / \nu = 0.33(1)$ for the RRN, and
$\beta / \nu = 0.44(2)$ for UCM networks, in reasonable agreement with
the theoretical values $\beta_\mathrm{th} / \nu_\mathrm{th} = 1/3$ and
$\beta_\mathrm{th} / \nu_\mathrm{th} = 0.40$, respectively, see
Figs.~\ref{fig:3}(c) and~\ref{fig:4}(c).

As in the case of percolation we can fit the position $\lambda_c(N)$ of
the peak of the susceptibility $\chi_3$ to formula analogous to
Eq.~(\ref{eq:31}) to obtain the values of the asymptotic critical point
$\lambda_c$ and of the exponent $\nu$.
\begin{figure}
  \centering
  \includegraphics[width=7cm]{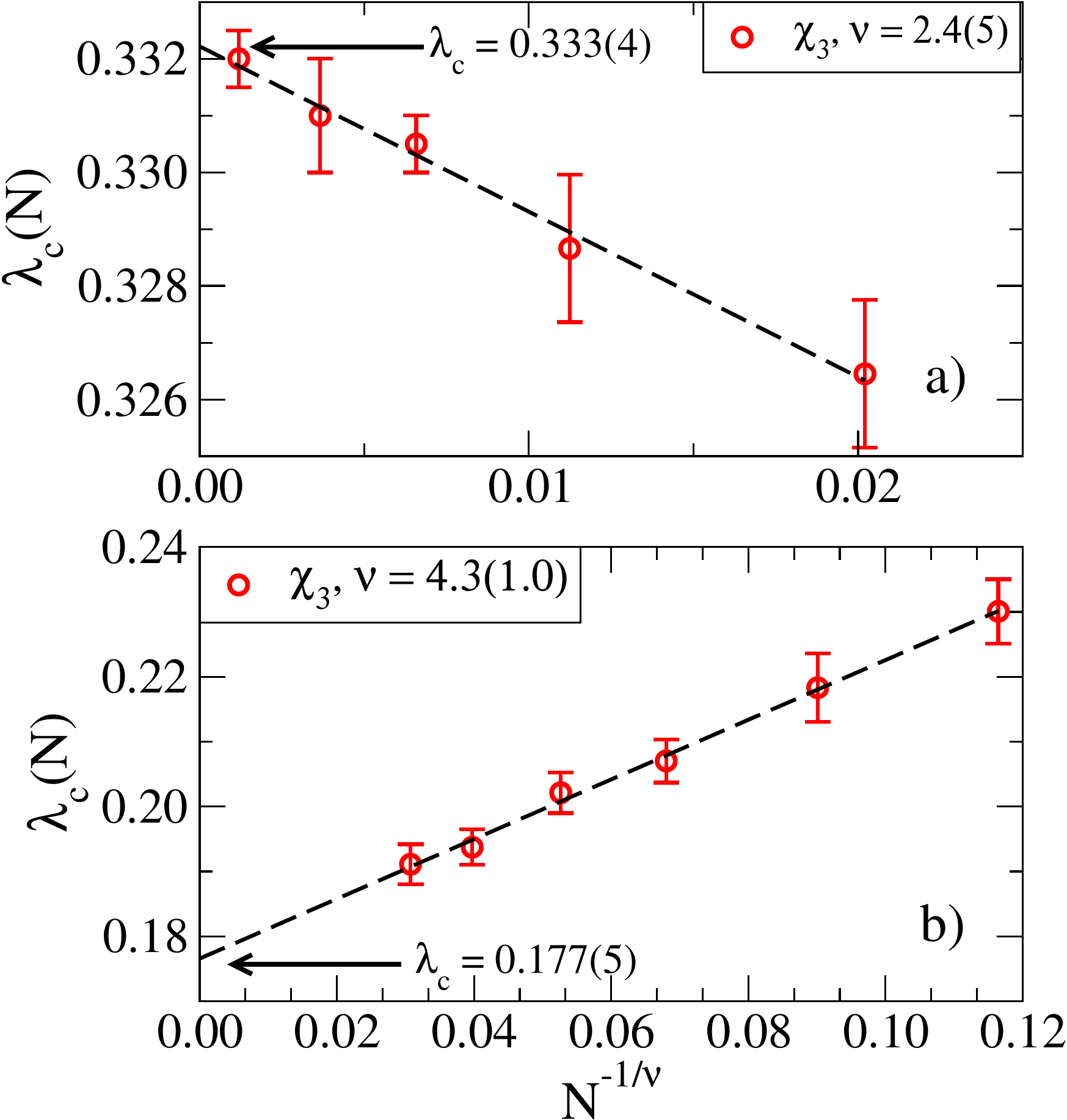}
  \caption{Plot of the peak position as a function of $N$ for the SIR
    dynamics on RRN (a) and UCM networks with degree exponent
    $\gamma_d = 3.5$ (b). The linear behaviors are in agreement with
    Eq.~(\ref{eq:31}).}
  \label{fig:pc_SIR}
\end{figure}
From such a non-linear fitting, see Fig.~\ref{fig:pc_SIR}, we obtain for
RRN the value $\nu=2.4(5)$ and a critical point
$\lambda_c=0.333(4)$. For UCM we obtain instead $\nu=4.3(1.0)$ and
$\lambda_c=0.177(5)$. In both cases there is a fair agreement with the
expected theoretical values.

\begin{figure}[t]
  \centering
  \includegraphics[width=\columnwidth]{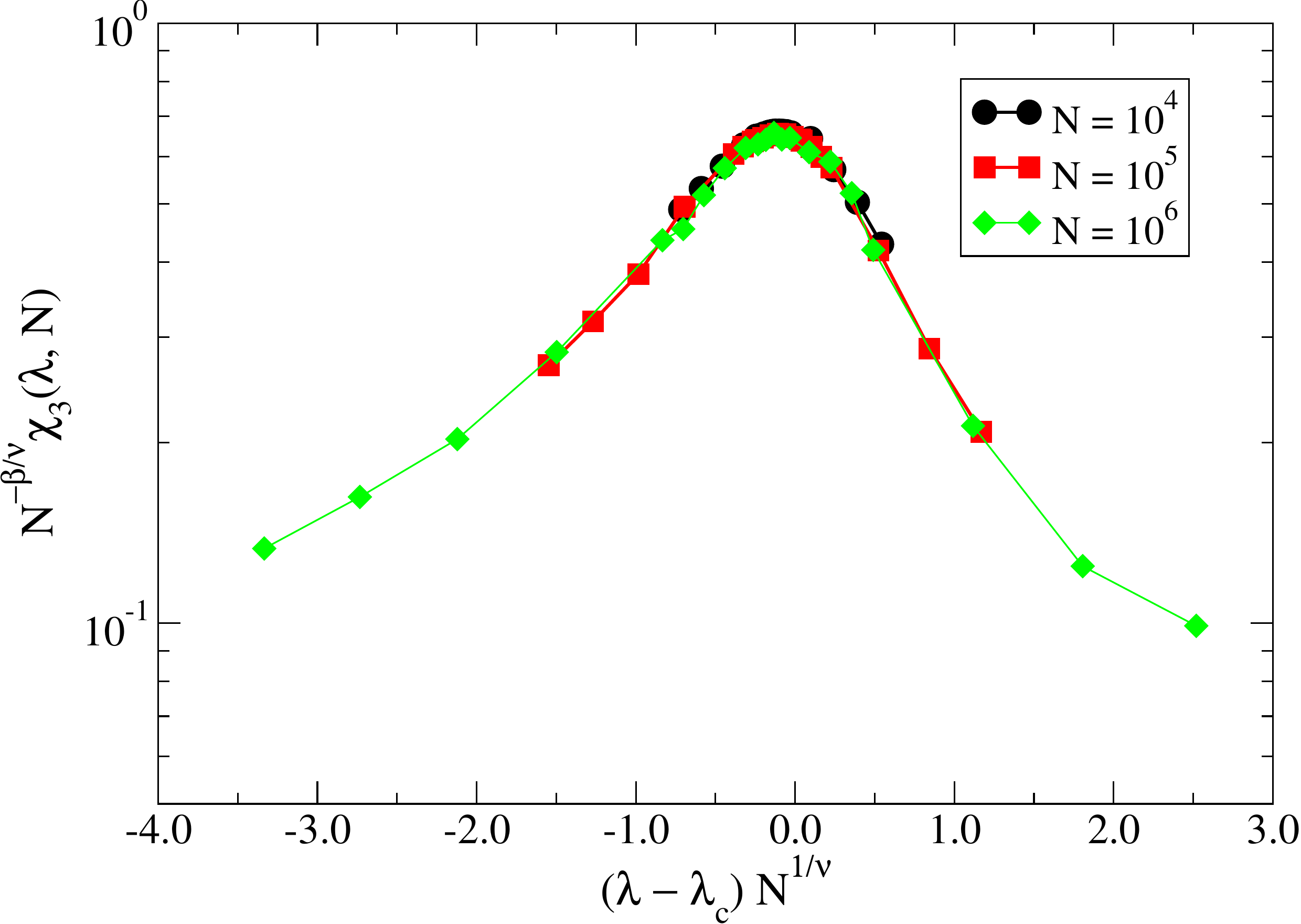}
  \caption{Data collapse analysis of the susceptibility $\chi_3$ for SIR
    model on RRN networks with degree $K=5$. We have used the exact
    theoretical values $\lambda_c^\mathrm{RRN, th} = 1/3$,
    $\nu_\mathrm{th}=3$, $\beta_\mathrm{th} / \nu_\mathrm{th} = 1/3$.}
  \label{fig:collapsSIRRNN}
\end{figure}
Performing a data collapse analysis, according to Eq.~(\ref{eq:2}), we
observe that data for RRN exhibits, as in the case of percolation, an
almost perfect collapse using the exact theoretical values
$\lambda_c^\mathrm{RRN, th} = 1/3$, $\nu_\mathrm{th} = 3$, and
$\beta_\mathrm{th} / \nu_\mathrm{th} = 1/3$, see
Fig.~\ref{fig:collapsSIRRNN}.
\begin{figure}[t]
  \centering
  \includegraphics[width=\columnwidth]{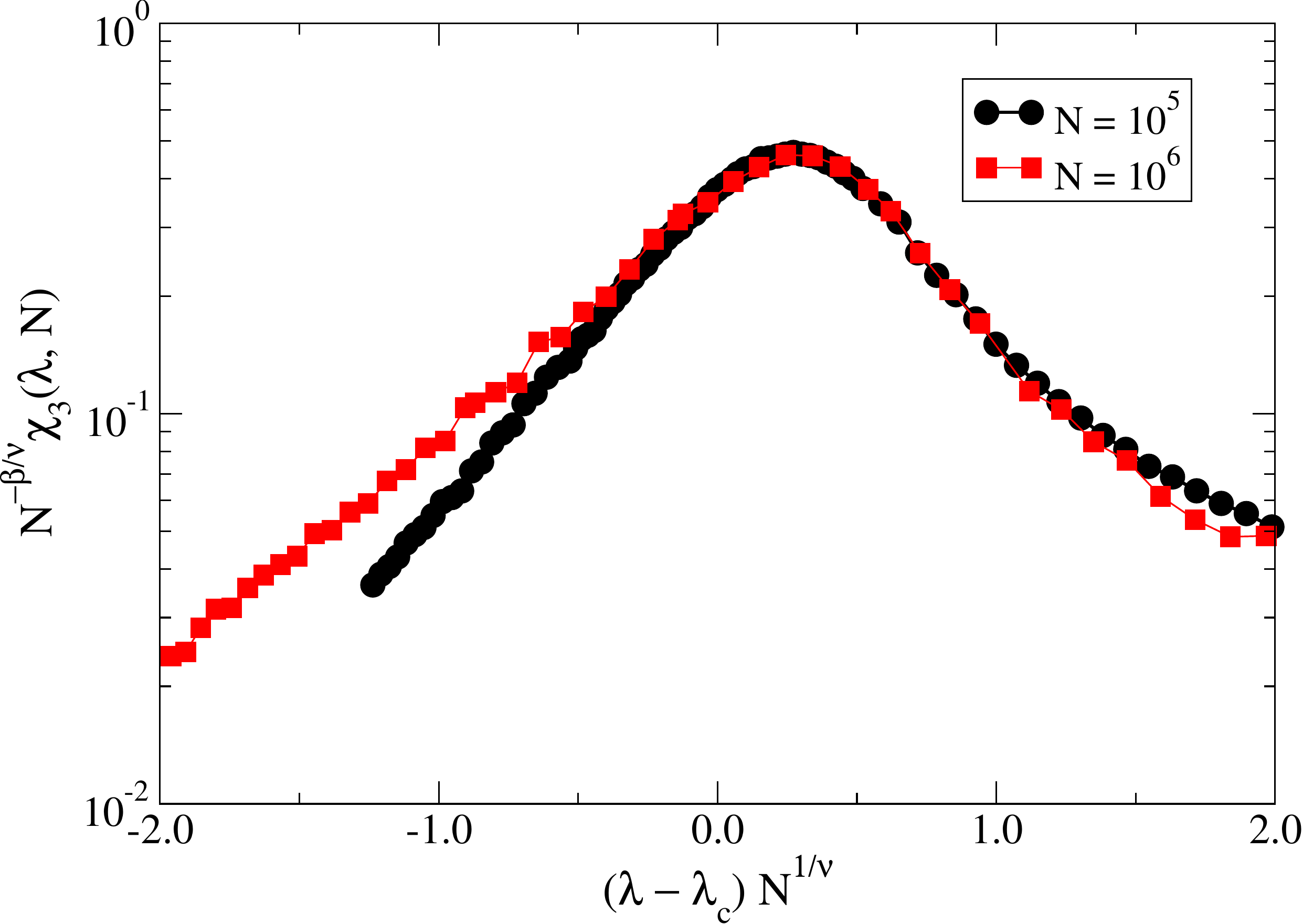}
  \caption{Data collapse analysis of the susceptibility $\chi_3$ for SIR
    model on UCM networks of degree exponent $\gamma=3.5$. We have used
    the numerical epidemic threshold $\lambda_c^\mathrm{UCM} = 0.185(5)$
    and the exponents $\nu = 4.3(5)$, $\beta/\nu = 0.43(3)$. }
  \label{fig:collapsSIRUCM}
\end{figure}
In the case of scale-free UCM networks, neither the theoretical
predictions nor the numerically estimated parameters provide a good
collapse of the susceptibilities $\chi_3$ for different network
sizes. This is again due to the uncertainties in the position of the
epidemic threshold, which are perceptible in the single network sample
data used of the collapse\footnote{Notice that we cannot average the
  data for the whole $\chi_3(p)$ since it would lead to a smoothing and
  rounding of the susceptibility peak.}. In this case, we proceed to
estimate the best collapse by minimizing the distance between the
rescaled plots using the Nelder-Mead unconstrained optimization
algorithm, as implemented in the Python package
\texttt{fssa}\footnote{Available at
  \texttt{http://pyfssa.readthedocs.org/en/stable/}.}. The best collapse
is obtained using the values $\lambda_c^\mathrm{UCM} = 0.185(5)$,
$\nu = 4.3(5)$ and $\beta/\nu = 0.43(3)$. The exponents are quite close
to the theoretical predictions. With respect to the value of the
numerical epidemic threshold, we can compare it with the numerical
critical point obtained for percolation by noticing that, from
Eqs.~(\ref{eq:12}) and~(\ref{eq:1}), we have
\begin{equation}
  \frac{1}{ \lambda_c} =  \frac{1}{ p_c} - 1.
  \label{eq:8}
\end{equation}
Using the numerical percolation value for UCM networks,
$p_c^\mathrm{UCM} \simeq 0.153$ in Eq.~(\ref{eq:8}), we obtain
$\lambda_c^\mathrm{UCM} \simeq 0.181$, in good agreement with the best
critical point from the data collapse analysis.

\begin{figure}
  \centering
  \includegraphics[width=\columnwidth]{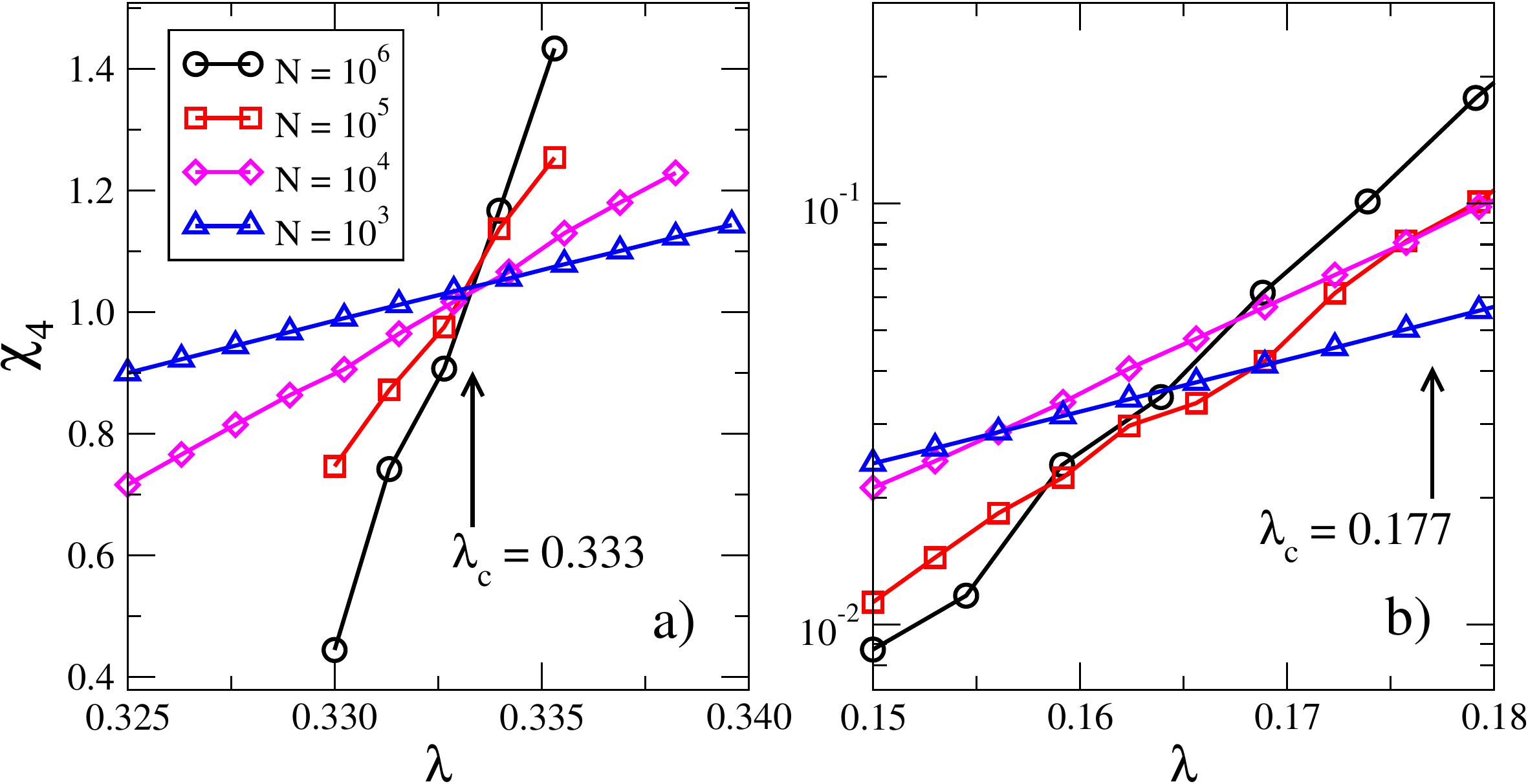}
  \caption{Close up of the quantity $\chi_4$ in the vicinity of the
    critical point for RRN networks (a) and UCM networks with
    $\gamma_d=3.5$ (b). In this last case, we consider the average over
    at least $10$ different network samples.}
  \label{fig:chi_4}
\end{figure}

Finally, in analogy with the case of percolation we consider also an
additional quantity, analogue to Binder's cumulant. In view of the
scalings in Eq.~(\ref{critscal}), the quantity
\begin{equation}
\chi_4 = \frac{\av{\overline{\phi}}^3}{\av{\overline{\phi^2}}^2} 
\end{equation}
should scale at criticality as $N^{1-(2\beta/\nu+\gamma/\nu)}$ and hence
be constant due to the hyperscaling relation Eq.~(\ref{eq:20}).  In
Fig.~\ref{fig:chi_4} we plot the behavior of this analogue of Binder's
cumulant for this dynamics in the vicinity of the critical point. For
RRN this quantity allows to determine with excellent precision the
location of the critical point as the intersection of the curves for
different size $N$.  In the case of UCM instead, the presence of large
sample-to-sample fluctuations spoils the determination of a single
intersection point.  In this case, even averaging the value of $\chi_4$
over several realizations does not lead to a reliable estimate of the
critical point.

\section{Discussion and conclusions}
\label{conclusions}

The numerical evaluation of epidemic thresholds in networks represents
an important issue, with practical implications in real world situations
\cite{Pastor-Satorras:2014aa}. Despite this fact, in the case of the SIR
model no clear prescription has been defined so far, and several
alternative approaches~\cite{Marrobook,Ferreira12,Colomer-de-Simon2014,PhysRevE.91.010801,Crapey2006,Shu15}
have been proposed and applied in the literature,
based on the use of a "susceptibility" measure, defined as a
quantity that, as a function of the spreading rate (in the SIR model) or
the occupation probability (for percolation), ought to show a maximum
located in the vicinity of the putative critical point, while decreasing
to a constant value away from it, in a similar fashion as the
susceptibility usually considered in equilibrium statistical mechanics
\cite{yeomans}. In the present paper we have performed a theoretical
analysis of different forms of susceptibilities that have been applied
to study the SIR model and the related percolation process. 
The analysis of three possible candidate susceptibilities indicates
that different forms of susceptibility are better suited to analyze
percolation or the SIR process. More specifically, the susceptibility
$\chi_3$, Eq.~(\ref{eq:17}), is the correct one for the SIR model, while the
susceptibilities $\chi_1$, Eq.~(\ref{eq:21}), and $\chi_2$,
Eq.~(\ref{eq:25}) are better suited for percolation, the latter
outperforming the former due to its fastest divergence at
criticality. This different performance is traced back to the different
nature of the numerical observables in SIR and percolation. While both
models can be exactly mapped one onto the other, different observables
can be measured for each of them in numerical simulations. So, while for the
percolation process one can easily extract the \textit{largest} cluster
of each percolation sample in order to define an order parameter, and
perform averages restricted over it, in the SIR process such distinction
is impossible, and one is forced to define an order parameter in terms
of averages over \textit{all} outbreaks (clusters), sampled
intrinsically with a probability proportional to their size.

\begin{figure}
  \centering
  \includegraphics[width=\columnwidth]{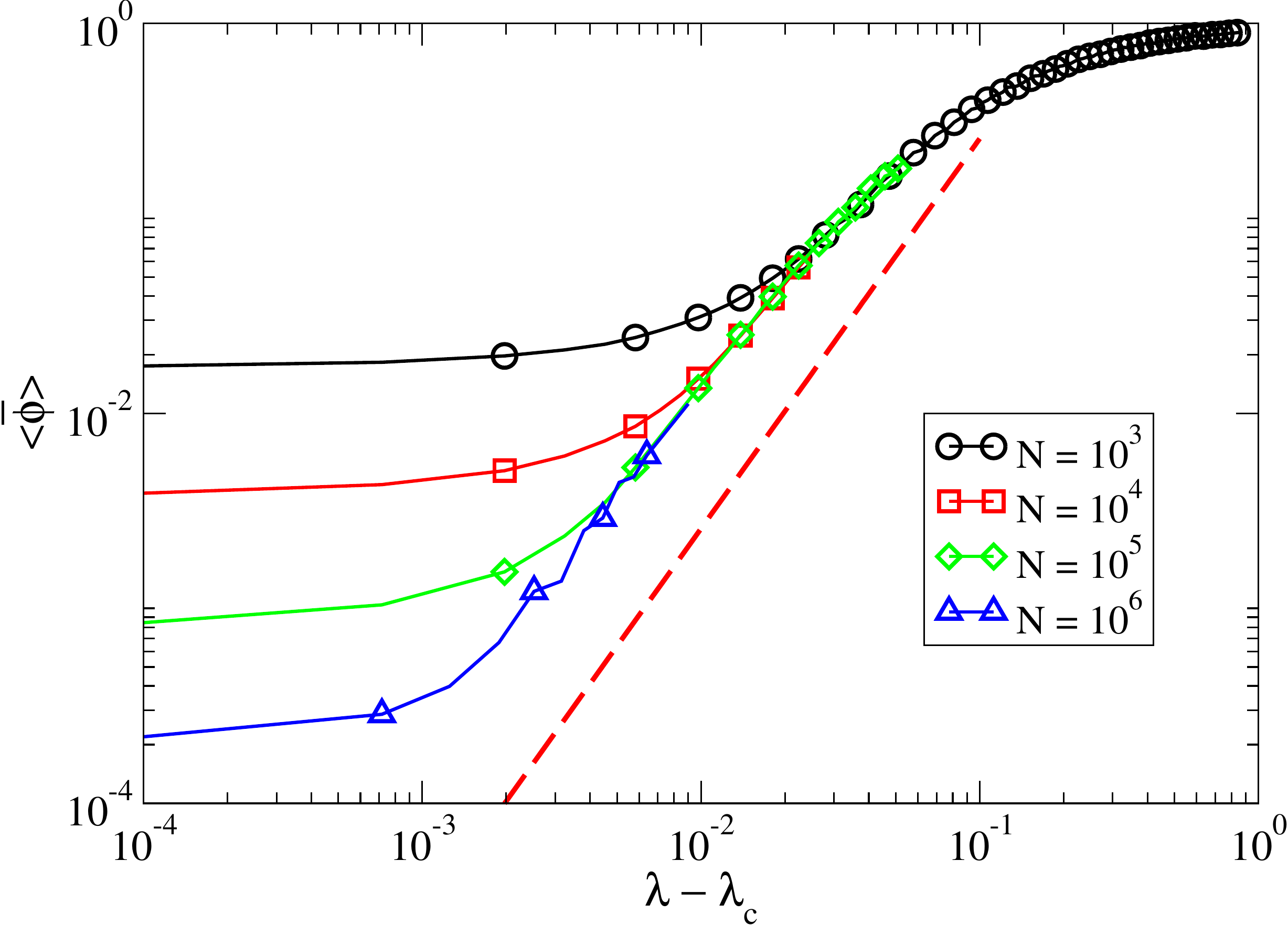}
  \caption{Order parameter $\av{ \overline{\phi}}$ (average outbreak
    size) for the SIR model on RRN with degree $K=10$ as a function of
    $\lambda-\lambda_c$. The dashed line represent the behavior
    $(\lambda-\lambda_c)^2$.}
  \label{fig:5}
\end{figure}

An additional consequence of the biased sampling of clusters in the SIR
model is an effect regarding the determination of the exponent $\beta$,
associated to the growth of the order parameter in the supercritical
phase.  When using the natural definition of the order parameter for the
SIR model, given by the average relative outbreak size,
$\av{ \overline{\phi}}$, from Eq.~(\ref{final}) we get, in the
supercritical phase
\begin{equation}
  \av{\overline{\phi}} = \frac{\av{G^2}}{N^2} \sim 
  \left[\lambda-\lambda_c \right]^{2 \beta}.
  \label{eq:29}
\end{equation}
As a consequence, if the order parameter $\av{\overline{\phi}}$ is plotted
versus $\lambda-\lambda_c$ an effective exponent $\beta_{SIR} = 2 \beta$
is found. This is confirmed in Fig.~\ref{fig:5}, where we present results
from simulations of the SIR process on RRN with fixed degree $K=10$,
for which $\beta=1$. This result by no means
invalidates the connection between SIR and percolation. It is only a
consequence of the unavoidable bias in the selection of percolation
clusters induced by the random choice of the initial seed of SIR
outbreaks. On the other hand, Fig.~\ref{fig:5} could potentially cast
some doubts on the validity of the HMF prediction for the exponent
$\beta$~\cite{refId0}, which coincides with the prediction in
Eq.~(\ref{eq:9}).  A closer scrutiny shows however that the HMF
prediction for SIR is correct: The order parameter considered in the HMF
theory is not $\av{\overline{\phi}}$, the average outbreak size, but
rather the probability that, at the end of the outbreak, a randomly
chosen node is recovered.  In the thermodynamic limit $N \to \infty$,
above the critical point, this quantity coincides with the relative size
of the giant component of the corresponding percolation problem.  This
explains why the critical exponent $\beta$ found by HMF theory for SIR
rightly coincides with the $\beta$ of bond percolation and is equal to a
half of the exponent found in SIR numerical simulations for the order
parameter $\av{\overline{\phi}}$ determined numerically.

The analysis presented here allows finally to reinterpret and clarify
some results appeared in the literature.  In Ref.~\cite{Shu15} it was
found numerically that epidemic variability $\chi'_3$ \cite{Crapey2006}
provides precise estimates of the SIR epidemic threshold, while using
$\chi_2$ leads to systematic errors.  The scaling analysis performed in
Sec.~\ref{sec:sir-model-analysis} allows to understand the reasons of
this observation. On the other hand Ref.~\cite{Lagorio2009755}  observed
that the average outbreak size does not correspond to the order
parameter in percolation, in agreement with the discussion above. The
authors of~\cite{Lagorio2009755} provide a numerical technique to make
these two quantities coincide: fix an outbreak size threshold $s_c$, and
perform averages only over outbreaks larger than this threshold. Again,
our results justify this recipe: The threshold introduced
biases the clusters averaged towards the theoretical largest cluster,
which is indeed the observable used to determine the order parameter in
percolation. 

\section*{Acknowledgments}

We thank Filippo Radicchi for a critical reading of the manuscript
and an anonymous reviewer for helpful comments.
R.P.-S. acknowledges financial support from the Spanish MINECO, under
project No.  FIS2013-47282-C2-2, EC FET-Proactive Project MULTIPLEX
(Grant No. 317532), and the ICREA Academia Foundation, funded by the
{\it Generalitat de Catalunya}.

\section*{Author Contribution Statement}
All authors contributed equally to the paper.

\bibliographystyle{epj}

\end{document}